%% file: main.tex
\documentclass[sigconf,authorversion,nonacm]{acmart}

\usepackage[]{hyperref}
\usepackage{xcolor,colortbl,xcolor-solarized}
\usepackage{cleveref}
\usepackage{listings}
\usepackage{subcaption}
\usepackage{soul}

\newcommand{\hlc}[2][yellow]{{%
    \colorlet{foo}{#1}%
    \sethlcolor{foo}\hl{#2}}%
}
\newcommand{\hlloop}{\hlc[green!20]}
\newcommand{\hlmem}{\hlc[blue!20]}
\newcommand{\hlcompute}{\hlc[red!20]}
\colorlet{lightgray}{gray!50}
\newcommand{\unchanged}{\textcolor{gray!50}}

\lstdefinestyle{base}{
    xleftmargin=3em,
    columns=fullflexible,
    numbers=left,         
    numberstyle=\small,   
    numbersep=2pt,        
    keywordstyle=\color{solarized-magenta},
    commentstyle=\color{solarized-blue},
    stringstyle=\color{solarized-orange},
    numberstyle=\color{solarized-base0},
    emphstyle=[1]\color{solarized-red},
    emphstyle=[2]\color{solarized-green},
    basicstyle={\scriptsize\linespread{0.8}\color{solarized-base03}\ttfamily}, showstringspaces=false,
    escapeinside={(*}{*)},
  }
\lstdefinestyle{tight}{
    basicstyle={\footnotesize\linespread{0.8}\color{solarized-base03}\ttfamily},
  }
\lstdefinestyle{cppstyle}{
  style=base,
  language=C++,
  emph=[1]{memref,index,idx,i32,f32,mref,stream,str,vec,vector,tu},
  emph=[2]{to,step,for,dae,dlc,dlcv,mem_str,alu_str,fwd_str,to_val,callback,done,pop_token,configure_tmu,pop_data,tkn,pop,push,buf_str},
}
\newcommand{\bnfdef}{\mathrel{::=}}
\newcommand{\bnfalt}{\mathrel{\mid}}

\usepackage{makecell}

\newcommand{\vcentered}[1]{\begingroup
\setbox0=\hbox{#1}\parbox{\dimexpr\ht0+\dp0}{\box0}\endgroup}

\begin{document}

\title{Ember: A Compiler for Efficient Embedding Operations on Decoupled Access-Execute Architectures}


\author{Marco Siracusa}
\email{marco.siracusa@bsc.es}
\orcid{0000-0003-2782-837X}
 \affiliation{
   \institution{Barcelona Supercomputing Center}
   \city{Barcelona}
   \country{Spain}
 }

\author{Olivia Hsu}
\email{owhsu@stanford.edu}
\orcid{0000-0002-4195-8106}
 \affiliation{
   \institution{Stanford University}
   \city{Stanford}
   \country{USA}
 }

\author{Victor Soria-Pardos}
\email{victor.soria@bsc.es}
\orcid{0000-0001-8337-6326}
 \affiliation{
   \institution{Barcelona Supercomputing Center}
   \city{Barcelona}
   \country{Spain}
 }

\author{Joshua Randall}
\email{joshua.randall@arm.com}
\orcid{0000-0002-5154-8688}
 \affiliation{
   \institution{Arm}
   \city{Austin}
   \country{USA}
 }

\author{Arnaud Grasset}
\email{arnaud.grasset@arm.com}
 \affiliation{
   \institution{Arm}
   \city{Biot}
   \country{France}
 }

\author{Eric Biscondi}
\email{eric.biscondi@arm.com}
 \affiliation{
   \institution{Arm}
   \city{Biot}
   \country{France}
 }

\author{Doug Joseph}
\email{doug.joseph@arm.com}
\orcid{0000-0001-7701-015X}
 \affiliation{
   \institution{Arm}
   \city{Austin}
   \country{USA}
 }

\author{Randy Allen}
\email{randy.allen@bsc.es}
 \affiliation{
   \institution{Barcelona Supercomputing Center}
   \city{Barcelona}
   \country{Spain}
 }
 
\author{Fredrik Kjolstad}
\email{kjolstad@stanford.edu}
\orcid{0000-0002-2267-903X}
 \affiliation{
   \institution{Stanford University}
   \city{Stanford}
   \country{USA}
 }

\author{Miquel Moretó Planas}
\email{miquel.moreto@bsc.es}
\orcid{0000-0002-9848-8758}
 \affiliation{
   \institution{Barcelona Supercomputing Center}
   \institution{Universitat Politècnica de Catalunya}
   \city{Barcelona}
   \country{Spain}
 }

\author{Adrià Armejach}
\email{adria.armejach@bsc.es}
\orcid{0000-0003-2869-668X}
 \affiliation{
   \institution{Barcelona Supercomputing Center}
   \institution{Universitat Politècnica de Catalunya}
   \city{Barcelona}
   \country{Spain}
 }

\renewcommand{\shortauthors}{Siracusa et al.}

\input{sec_00_abstract}

\maketitle

\input{sec_01_introduction}
\input{sec_02_architectural_challenges}
\input{sec_03_dae_potential}
\input{sec_04_dae_abstraction}
\input{sec_05_ember_overview}
\input{sec_06_slc_ir}
\input{sec_07_optimizations}
\input{sec_08_evaluation}

\input{sec_09_related}
\input{sec_10_conclusions}

\bibliographystyle{ACM-Reference-Format}
\bibliography{refs}

\end{document}

%% file: sec_00_abstract.tex
\begin{abstract}
Irregular embedding lookups are a critical bottleneck in recommender models, sparse large language models, and graph learning models. In this paper, we first demonstrate that, by offloading these lookups to specialized access units, Decoupled Access-Execute (DAE) processors achieve 2.6$\times$ higher performance and 6.4$\times$ higher performance/watt than GPUs on end-to-end models. Then, we propose the Ember compiler for automatically generating optimized DAE code from PyTorch and TensorFlow. Conversely from other DAE compilers, Ember features multiple intermediate representations specifically designed for different optimization levels. In this way, Ember can implement all optimizations to match the performance of hand-written code, unlocking the full potential of DAE architectures at scale.
\end{abstract}

%% file: sec_01_introduction.tex
\section{Introduction}
\label{sec:intro}

Recommender models, sparse large language models, and graph-learning models represent categorical features---such as users, products, and words---using dense embedding vectors~\cite{glove2014emnlp}. Due to the vast number of categories, interactions between these features are inherently sparse~\cite{recnmp2020isca}. For instance, recommender models must lookup the embedding vectors of the products a user has interacted with, often loading hundreds of vectors among millions~\cite{dlrm2020hpca}. This leads to irregular memory accesses with low cache reuse~\cite{caches2022memsys}, posing significant challenges for traditional architectures~\cite{mtia2023isca,bigbird2024neurips}.

In this paper, we first demonstrate that, by offloading embedding lookup to specialized access units, Decoupled Access-Execute (DAE) processors achieve 2.6$\times$ higher performance and 6.4$\times$ higher perf/watt than GPUs on end-to-end models. Then, we design the Ember compiler to automatically lower PyTorch~\cite{pytorch2019neurips} and TensorFlow~\cite{tf2015whitepaper} embedding operations to high-performance DAE code.

The goal of a DAE compiler is to decouple memory accesses from computation and generate optimized code for the access unit and execute unit. However, optimizing already decoupled code is challenging since memory access is separated from computation and (de)serialized through queues. This breaks the control flow and data flow of the input program, hindering global optimizations such as vectorization or code motion across access and execute unit.

For this reason, Ember features multiple Intermediate Representations (IRs), each one designed for a different optimization level. Ember's low-level Decoupled Lookup-Compute (DLC) IR is designed to abstract implementation details of the underlying DAE architecture and facilitate local, target-agnostic optimizations of lookup code and compute code separately. Ember's high-level Structured Lookup-Compute (SLC) IR is designed to also abstract the queue (de)serialization mechanism and facilitate global optimizations within lookup and compute code. By lowering PyTorch and TensorFlow embedding operations through the SLC and DLC IRs, Ember can implement all necessary optimizations to match the performance of hand-optimized code, enabling the full DAE potential at no programmability cost.

Overall, our contributions are:
\begin{enumerate}
    \item A characterization of a large class of embedding operations in machine learning models to study the fundamental scaling limitations of traditional architectures (\Cref{sec:architectural-challenges}).
    \item An evaluation of the potential of DAE architectures for embedding operations (\Cref{sec:dae-potential}).
    \item The DLC IR, a low-level IR designed to abstract and optimize decoupled lookup and compute code (\Cref{sec:dae-abstraction}).
    \item An end-to-end implementation of the Ember compiler to lower PyTorch/TensorFlow embedding operations to such abstraction, and then to DAE architectures (\Cref{sec:ember-overview}).
    \item The SLC IR, a high-level IR designed for global optimizations of DAE embedding operations, and algorithms to lower to/from it (\Cref{sec:decoupling}).
    \item Optimizations for DAE embedding operations (\Cref{sec:optimizations}).
\end{enumerate}
After introducing these contributions, this paper evaluates Ember (\Cref{sec:evaluation}) and discusses its impact in the field (\Cref{sec:related-work}).

%% file: sec_02_architectural_challenges.tex
\section{Challenges of Embedding Operations} \label{sec:architectural-challenges}

\begin{figure}
    \centering
    \includegraphics[width=\linewidth]{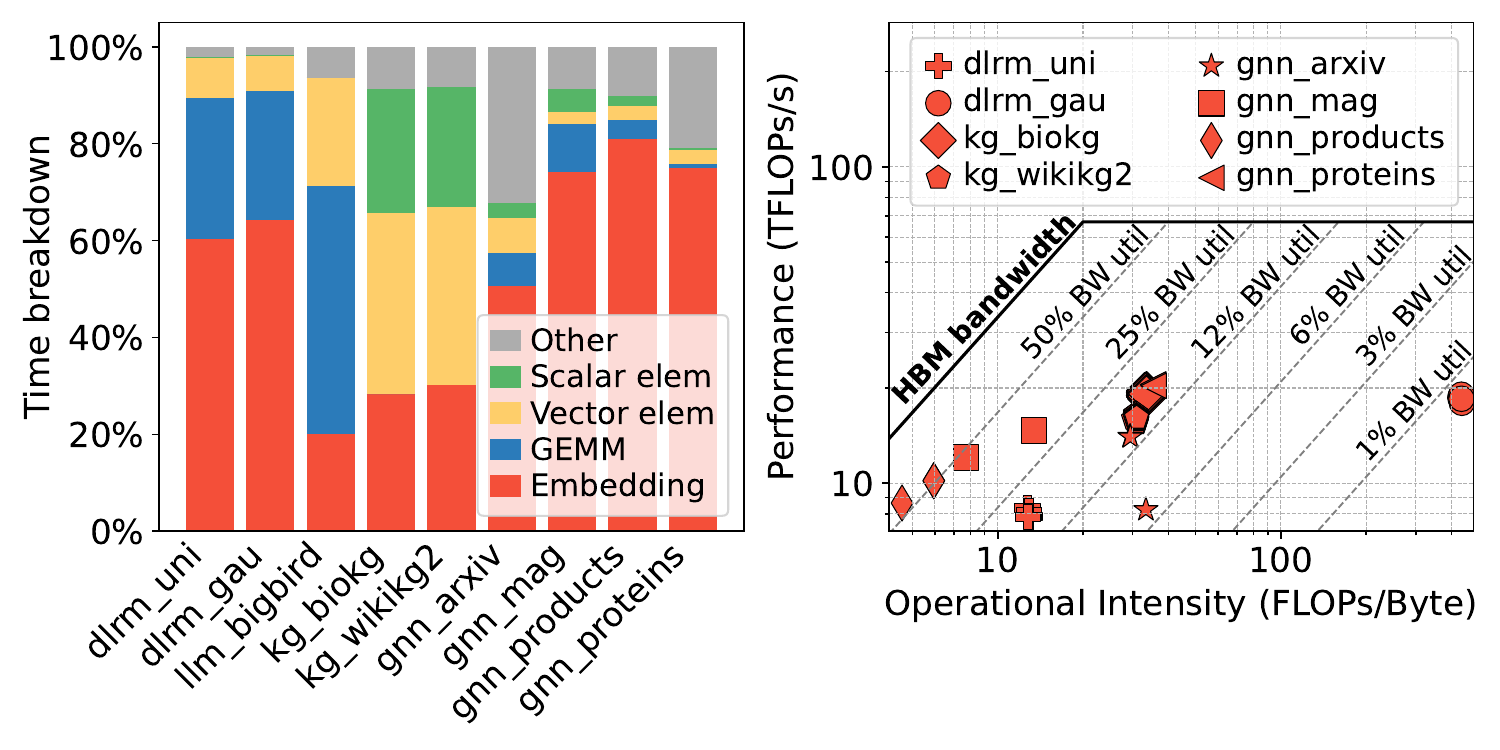}
    \vspace{-2.5em}
    \caption{Deep-learning recommendation models (\texttt{dlrm}) (\Cref{sec:dlrms}), large language models (\texttt{llm}) with sparse attention (\Cref{sec:llms}), knowledge graphs (\texttt{kg}) (\Cref{sec:graph-learning}), and graph neural networks (\texttt{gnn}) (\Cref{sec:graph-learning}) heavily rely on embedding operations that do not perform efficiently even on modern Nvidia H100 GPUs~\cite{nvidiah100}. All experiments use highly-optimized models from the literature (\Cref{sec:characterization-implications}).}
    \label{fig:gpu-benchmark}
\end{figure}

\Cref{fig:gpu-benchmark} shows that several machine learning models heavily rely on embedding operations that achieve low system utilization even on the latest datacenter GPUs. This section examines the characteristics of embedding operations, their architectural implications, and the challenges they present for performance scaling.

\subsection{Background} \label{sec:background}

\begin{figure}
    \centering
    \includegraphics[width=\linewidth]{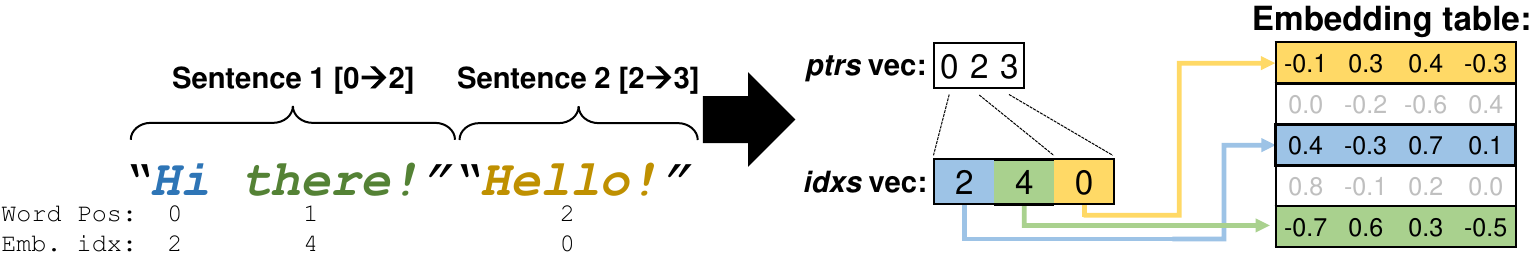}
    \vspace{-2em}
    \caption{Feature embedding requires scattered memory lookups to fetch embedding vectors from embedding tables.}
    \label{fig:irregular-lookup}
    \vspace{-0.5em}
\end{figure}

Modern machine learning models process an increasing amount of \emph{categorical features}~\cite{caches2022memsys}---representation variables that can take on a distinct set of values and are not comparable. Example categorical features include product categories, words, and actors in a movie. As these categories have no numerical meaning, they cannot be directly processed with deep neural networks~\cite{dlrm2020hpca}. Hence, machine learning models embed the numerical meaning of these categories into dense \emph{embedding vectors}~\cite{glove2014emnlp}, which are all stored into large \emph{embedding tables}. \Cref{fig:irregular-lookup} illustrates how categorical features are processed during inference. Incoming categories are tokenized into IDs, which are used to \emph{lookup} into the embedding table. Embedding lookup is generally fused with downstream computation (e.g., reduction) in an optimized \emph{embedding operation}.

\subsection{Characterization and Implications} \label{sec:characterization-implications}

\input{tab_characterization}

\Cref{tab:characterization} characterizes embedding operations across various machine learning models. For each model, we first analyze its loop hierarchy and compute-per-lookup ratio, which proxy compute requirements. Then, we compute the memory footprint of their embedding tables to understand their storage demand. Finally, we characterize their spatio-temporal locality to gauge the effectiveness of caches and prefetchers for embedding lookups. We characterize temporal locality through the concept of reuse distance---the number of other vectors accessed before a vector is accessed again~\cite{recshard2022asplos}. If a cache can hold $x$ vectors, an access with reuse distance greater than $x$ will likely result in eviction. For a given input, we first construct a histogram of all reuse distances, which we then integrate and normalize to obtain the Cumulative Distribution Function ($CDF$). Hence, $CDF(x)$ proxies the hit probability of that cache. The spatial locality, instead, is characterized by the size of the embedding vector.

\subsubsection{Deep Learning Recommendation Models (DLRMs)} \label{sec:dlrms}

DLRMs recommend content to users by processing a large set of categorical features under strict latency constraints~\cite{dlrm2020hpca,caches2022memsys}. DLRMs currently account for a large fraction of inference cycles in production data centers, with embedding operations consuming a significant runtime portion~\cite{dlrm2020hpca}, as shown in \Cref{fig:gpu-benchmark}. 

Categorical features in DLRMs are generally multi-hot, meaning they can take on multiple values (e.g. movie genres or cast). At inference time, the embedding vectors corresponding to these values need to be looked up and aggregated with, for instance, an element sum. High-performance implementations (1)~fuse these two operations to avoid materializing intermediate results and (2)~batch together the categories of multiple queries to improve system throughput (column 2 in \Cref{tab:characterization}), which yields a compute-per-lookup ratio of 1 (column 3 in \Cref{tab:characterization}). The \texttt{nn.EmbeddingBag} (EB) in PyTorch~\cite{pytorch2019neurips} and Sparse-Lenghts Sum (SLS) function in Caffe2~\cite{caffe2014arxiv} implement these optimizations.

DLRMs invoke EB/SLS functions 10s---1000s of times to aggregate all of the categorical features in a batch~\cite{caches2022memsys} (clusters of crosses and circles in \Cref{fig:gpu-benchmark}). Each invocation aggregates 10s---100s embedding vectors from a single table with millions of entries~\cite{caches2022memsys}, each vector storing 32-256 elements (column 6 in \Cref{tab:characterization}). While this necessitates a large memory footprint (column 4 in \Cref{tab:characterization}), embedding vectors are accessed with some degree of reuse~\cite{recshard2022asplos,mixed2021press,neural2020acm,understanding2021hpca}, and modern caches can filter a significant fraction of embedding accesses~\cite{caches2022memsys}. If we assume embedding vectors with 256 FP32 elements, a 1MB cache can fit 1K vectors. Given the CDF of a real-world input such as the Criteo 1TB dataset~\cite{criteo2024} (column 5 in \Cref{tab:characterization}), \texttt{CDF\_ftr0}(1K)=63\% whereas \texttt{CDF\_ftr2}(1K)=99\%, which is the cache hit probability. Similarly, a 2MB cache would filter 65---99\% of the total accesses. Hence, while caches are essential for DLRMs, they bring diminishing returns. Moreover, given the relatively small embedding size, prefetchers would yield limited benefits~\cite{recnmp2020isca}.

\Cref{fig:gpu-benchmark} tests two high-performance DLRMs~\cite{naumov2019dlrm}. While \texttt{dlrm\_rnd} has a random access distribution, \texttt{dlrm\_uni} has a CDF similar to \texttt{criteo\_ftr1} in \Cref{tab:characterization}. In both cases, many accesses cannot be filtered by the GPU cache, and require hundreds of cycles to reach memory. As discussed in \Cref{sec:scaling-challenges}, GPUs cannot issue enough requests to hide such latency, only achieving 28\% utilization.

\subsubsection{Sparse Attention Mechanisms} \label{sec:llms}

Large Language Models (LLMs) embed text into vectors~\cite{bert2019acl}. Transformers such as Google's BigBird~\cite{bigbird2024neurips} sparsify the attention mechanism to only compare a selected subset of input tokens and improve memory footprint and performance on long input sequences. Although this technique drastically reduces inference latency, it requires an extra step to gather embedding blocks. This gathering step is generally not fused with downstream deep neural networks~\cite{bigbird2024neurips}, requiring only embedding lookups with no compute (SpAttn in \Cref{tab:characterization} column 3). The execution time of this gather operation depends on many factors, and can take up to 20\% if gathering, for instance, 8 random blocks per query element, as shown in \Cref{fig:gpu-benchmark}.

This operation differs from a normal \texttt{tf.gather}~\cite{tf2015whitepaper} as it replicates blocks of embeddings from the key tensor into the query tensor. As shown in \Cref{tab:characterization}, the larger the block, the larger the spatio-temporal locality over keys (higher horizontal CDF part). However, the smaller the block, the lower the compute required. Hence, larger blocks enable better system utilization on flop-oriented machines, and smaller blocks enable lower inference latency on machines with more efficient memory subsystems.

\subsubsection{Graph Machine Learning Models} \label{sec:graph-learning}

\input{tab_graph_inputs}

Graph machine learning models such as Graph Neural Networks (GNNs)~\cite{ogb2020neurips}, Message Passing (MP) models~\cite{fusedmm2021ipdps}, and Knowledge Graphs (KGs)~\cite{bernerslee2001semanticweb} embed node or edge features of a graph into vectors. GNNs, for instance, perform inference by chaining layers of deep neural networks and graph convolutions, which gather embeddings from neighbors with the SpMM-like operation shown in \Cref{tab:characterization}. These SpMM-like embedding operations generally have a higher compute-per-lookup ratio than DLRMs, with MP having the highest. \Cref{tab:graph-inputs} shows typical inputs for these models, with their CDFs reported in \Cref{tab:characterization}. Overall, graph-learning models often have a lower temporal locality (flatter CDF) than DLRMs and LLMs, resulting in more memory accesses. Hence, even \texttt{gnn\_proteins}, which has the highest reuse among GNNs, only achieves 32\% peak GPU performance.

\subsection{Challenges on Performance Scaling} \label{sec:scaling-challenges}

\input{fig_embedding_implications}

\begin{figure}
    \centering
    \includegraphics[width=\linewidth]{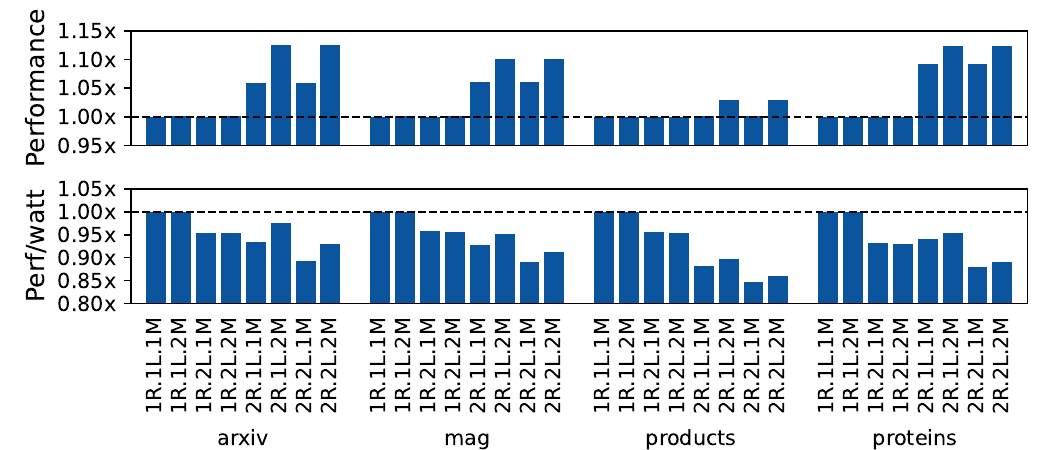}
    \vspace{-1em}
    \caption{Implications of scaling up the memory-level parallelism of a traditional CPU core (details in \Cref{fig:dae-specs}) for GNNs embedding operations (\Cref{sec:graph-learning}). Doubling reorder buffer, load-store queue, and L1D miss-status handling registers (\texttt{2R.2L.2M}) provides limited performance improvements and worse perf/watt than off-the-shelf cores (\texttt{1R.1L.1M}).}
    \label{fig:dse-cpu}
\end{figure}

Compute cores in traditional architectures---such as CPUs and GPUs---are optimized to access data in local caches, which are populated by CPU prefetchers or GPU DMAs~\cite{nvidiah100}. However, these mechanisms are ineffective for irregular workloads such as embedding operations~\cite{tmu2023micro}. As shown in \Cref{fig:ltu-histogram}, this results in a high volume of long memory accesses, where up to 86\% of requests are more than 10$\times$ longer than L1D accesses, and up to 36\% are more than 100$\times$ longer. However, compute units such as GPU SMs and CPU cores are not dimensioned for such long memory access latencies~\cite{fpga2022ccpe}. For instance, as shown in \Cref{fig:outstanding-requests} and \ref{fig:request-throughput}, off-the-shelf CPU cores can only track a few of these long memory requests, achieving low throughput (i.e. loads/cycle) and performance.

Increasing the memory-level parallelism of a CPU core to track more in-flight requests requires scaling up the reorder buffer, load-store queue, and miss-status handling registers~\cite{spchar2024jpdc}. However, these resources scale inefficiently. As shown in \Cref{fig:dse-cpu}, doubling them only improves performance by up to 12\%, with a 21\% power overhead. Besides being inefficient, scaling up these resources introduces challenges in timing closure~\cite{cad2022tc}, which would require lowering the core's frequency and, consequently, compute performance.

Scaling out the number of cores is also ineffective. As shown in \Cref{fig:hbm-bw}, we would need 43---72 traditional CPU cores to saturate a single HBM2 stack. Hence, we would need an unfeasible amount of cores to saturate the large memory bandwidth of current architectures~\cite{mess2024micro}. Similarly, as shown in \Cref{fig:gpu-benchmark}, current GPUs utilize 0.08\%---52\% of the HBM bandwidth. To achieve full HBM utilization, GPUs should use 2$\times$---12$\times$ more warps, which is challenging. Hence, alternative solutions are needed.

%% file: tab_characterization.tex
\begin{table*}[]
    \centering
    \footnotesize
    \begin{tabular}{clcccc}
        \toprule
        \makecell[c]{\textbf{Embedding Operation} \\ \textbf{and Model Class}} & \makecell[l]{\textbf{Embedding operation} \\ \textbf{reference implementation}} & $\frac{\textbf{N}_\textbf{compute}}{\textbf{N}_\textbf{lookups}}$ & \textbf{Memory footprint} & \makecell[c]{\textbf{Temporal locality} \\ \textbf{(reuse distance CDF)}} & \makecell[c]{\textbf{Spatial locality} \\ \textbf{(vector dim.)}}\\
        \toprule
        \makecell[c]{EmbeddingBag (EB) or \\ Sparse-Lenghts Sum \\ (SLS) for Deep-Learning \\ Recommendation Models \\ (DLRMs) (\Cref{sec:dlrms})} & \makecell[l]{$\forall$ segment in batch\\ \hspace{0.5em}$\forall$ category in segment\\ \hspace{1.0em}\textbf{lookup} vector\\ \hspace{1.0em}\textbf{accumulate} vector} & \normalsize$\frac{1}{1}$ & \makecell{All embedding tables \\ require GBs~\cite{dlrm2020hpca}\\ to TBs~\cite{caches2022memsys}} & \vcentered{\includegraphics[width=.23\linewidth]{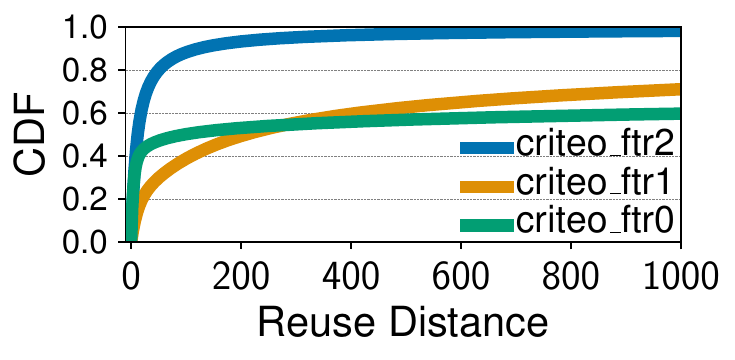}} & 32-256\\
        \hline
        \makecell[c]{Sparse attention (SpAttn) \\ for Large Language Models \\ (LLMs) (\Cref{sec:llms})} & \makecell[l]{$\forall$ block in Q input\\ \hspace{0.5em}$\forall$ block in K input\\ \hspace{1.0em}$\forall$ token in K block\\ \hspace{1.0em}\textbf{lookup} K vector\\ \hspace{1.5em}$\forall$ token in Q block\\ \hspace{2em}\textbf{copy} K vector} & \normalsize$0$ & \makecell{K-V tensors are MBs \\ Out tensor is MB-GB} & \vcentered{\includegraphics[width=.23\linewidth]{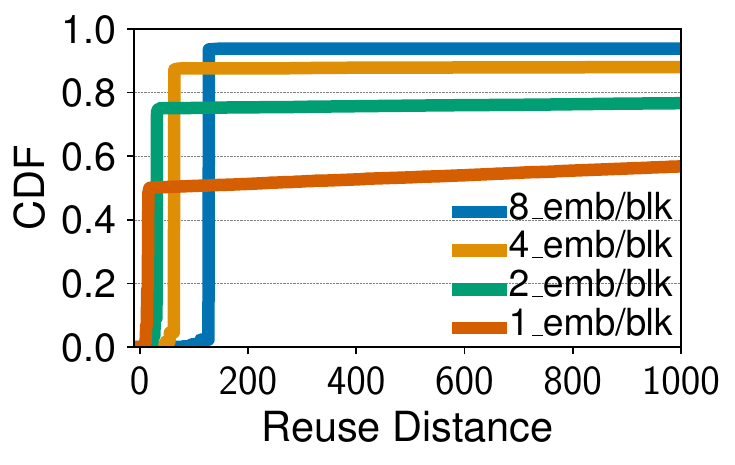}} & $\leq$ 8x8x512\\
        \hline
        \makecell[r]{Sparse Matrix-Matrix \\ Multiplication (SpMM) for \\ Graph Neural Networks \\ (GNNs) (\Cref{sec:graph-learning})} & \makecell[l]{$\forall$ vertex in graph\\ \hspace{0.5em}$\forall$ neighbor of vertex\\ \hspace{1em}\textbf{lookup} vector\\ \hspace{1em}\textbf{rescale} and \textbf{accumulate} vector} & \normalsize$\frac{2}{1}$ & \makecell{Single embedding table \\ GBs large} & \vcentered{\includegraphics[width=.23\linewidth]{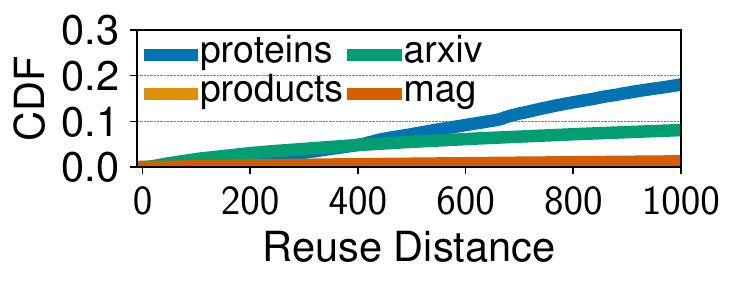}} & 8-349\\ 
        \hline
        \makecell[c]{Message Passing (MP) \\ models (\Cref{sec:graph-learning})} & \makecell[l]{$\forall$ vertex in graph\\ \hspace{1em}\textbf{lookup} vertex vector\\ \hspace{1em}$\forall$ neighbor of vertex\\ \hspace{2em}\textbf{lookup} neighbor vector\\ \hspace{2em}\textbf{multiply} vertex-neighbor vectors\\ \hspace{2em}\textbf{scale} resulting vector\\ \hspace{2em}\textbf{reduce} into tmp\\ \hspace{1em}\textbf{multiply} tmp by vertex vector\\ \hspace{1em}\textbf{accumulate} resulting vector} & \normalsize$\leq\frac{5}{2}$ & \makecell{Two embedding tables, \\ each GBs large} & \vcentered{\includegraphics[width=.23\linewidth]{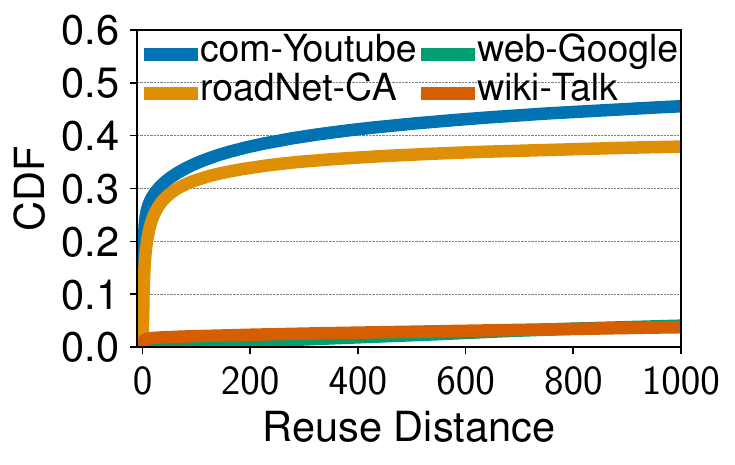}} & 128\\
        \hline
        \makecell[c]{Knowledge Graphs \\ (KGs) (\Cref{sec:graph-learning})} & \makecell[l]{$\forall$ sample in batch\\ \hspace{1em}\textbf{lookup} vector\\ \hspace{1em}\textbf{compute} head-rel-tail norm} & \normalsize$\frac{4}{3}$ & \makecell{Single embedding table \\ GBs large} & \vcentered{\includegraphics[width=.23\linewidth]{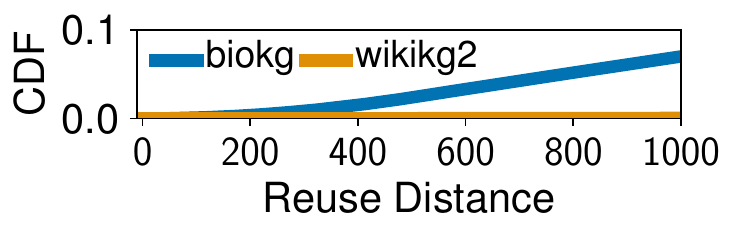}} & 512\\
        \bottomrule
    \end{tabular}
    \caption{Characterization of the embedding operations in \Cref{sec:characterization-implications}. The compute-per-lookup ratio captures code complexity. The memory footprint defines storage requirements for all embedding vectors. The temporal locality proxies cache efficiency through the Cumulative Distribution Function ($CDF$) of the vector reuse distance. The spatial locality proxies prefetcher efficiency through the number of embedding elements in each vector.}
    \label{tab:characterization}
    \vspace{-2em}
\end{table*}

%% file: tab_graph_inputs.tex
\begin{table}
  \centering
  \footnotesize
\[
\begin{tabular}{cccccc}
\toprule
  \textbf{Model} & \textbf{Input} & \textbf{\# Nodes} & \textbf{\# Edges} & \textbf{Layers sizes} \\
\toprule
  GNN & \texttt{arxiv}~\cite{ogb2020neurips} & 0.2M & 1.2M & 128 $\times$ 256 $\times$ 256 $\times$ 40 \\
  GNN & \texttt{mag}~\cite{ogb2020neurips} & 1.9M & 21.1M & 128 $\times$ 256 $\times$ 349 \\
  GNN & \texttt{products}~\cite{ogb2020neurips} & 2.4M & 61.9M & 100 $\times$ 256 $\times$ 256 $\times$ 47 \\
  GNN & \texttt{proteins}~\cite{ogb2020neurips} & 0.1M & 39.6M & 8 $\times$ 256 $\times$ 256 $\times$ 112\\
\hline
  MP & \texttt{com-Youtube}~\cite{snapnets2014} & 1.1M & 6.0M & 128 \\
  MP & \texttt{roadNet-CA}~\cite{snapnets2014} & 2.0M & 5.5M & 128 \\
  MP & \texttt{web-Google}~\cite{snapnets2014} & 0.9M & 5.1M & 128 \\
  MP & \texttt{wiki-Talk}~\cite{snapnets2014} & 2.4M & 5.0M & 128 \\
\hline
  KG & \texttt{biokg}~\cite{ogb2020neurips} & 0.1M & 5.1M & 512 \\
  KG & \texttt{wikikg2}~\cite{ogb2020neurips} & 2.5M & 17.1M & 512 \\
\bottomrule
\end{tabular}
\]
  \caption{Typical inputs for graph-learning models.}
  \label{tab:graph-inputs}
  \vspace{-2em}
\end{table}

%% file: fig_embedding_implications.tex
\begin{figure}
    \centering
    \includegraphics[width=\linewidth]{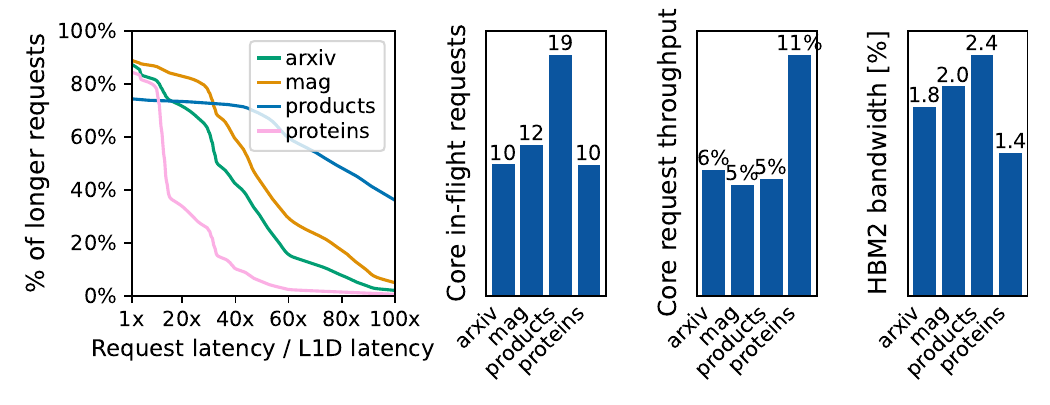}
    \vspace{-1em}
    \begin{subfigure}[b]{0.4\linewidth} 
        \vspace{-4em}
        \caption{}
        \label{fig:ltu-histogram}
    \end{subfigure}
    \begin{subfigure}[b]{0.19\linewidth} 
        \vspace{-4em}
        \caption{}
        \label{fig:outstanding-requests}
    \end{subfigure}
    \begin{subfigure}[b]{0.19\linewidth} 
        \vspace{-4em}
        \caption{}
        \label{fig:request-throughput}
    \end{subfigure}
    \begin{subfigure}[b]{0.19\linewidth} 
        \vspace{-4em}
        \caption{}
        \label{fig:hbm-bw}
    \end{subfigure}
    \caption{Architectural implications of embedding lookups on a traditional CPU core (details in \Cref{fig:dae-specs}). A large fraction of embedding lookups in GNNs models (\Cref{sec:graph-learning}) take orders of magnitude longer than L1D accesses (a). For instance, more than 74\% of \texttt{product}'s lookups are 10$\times$ longer than an L1D access, and 40\% more than 100$\times$ longer. However, traditional CPU cores have limited memory-level parallelism, and can only track a few in-flight lookups (b), stalling the CPU pipeline. This results in low memory request throughput (c), and low HBM per core utilization (d).}
    \label{fig:embedding-implications}
\end{figure}

%% file: sec_03_dae_potential.tex
\section{The Potential of DAE Architectures} \label{sec:dae-potential}

In this section, we demonstrate that Decoupled Access-Execute (DAE) multicore processors that offload embedding lookup to specialized units outperform GPUs in performance and perf/watt.

\subsection{Target DAE Architecture} \label{sec:dae-architecture} 

\begin{figure}
\includegraphics[width=1\linewidth]{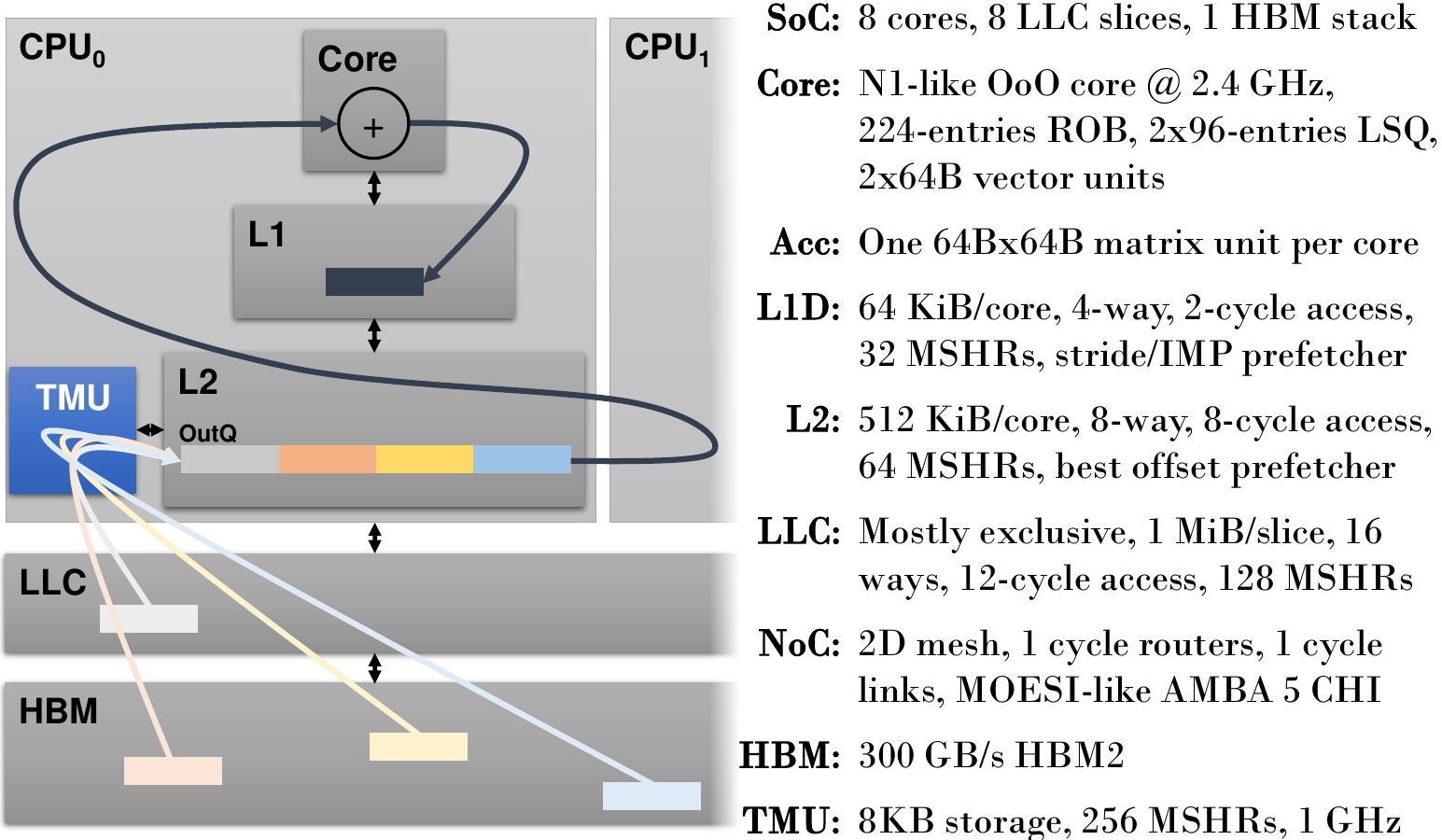} 
    \begin{subfigure}[b]{0.48\linewidth} 
        \caption{}
        \label{fig:dae-diagram}
    \end{subfigure}
    \begin{subfigure}[b]{0.48\linewidth} 
        \caption{}
        \label{fig:dae-specs}
    \end{subfigure}
\vspace{-1.5em}
\caption{A DAE processor. Each traditional core offloads embedding lookup to an access unit like the TMU~\cite{tmu2023micro}.}
\label{fig:dae-processor}
\end{figure}

\Cref{fig:dae-processor} shows the target DAE processor for this study, a traditional multicore processor that features one Tensor Marshaling Unit (TMU)~\cite{tmu2023micro} per traditional core to offload embedding lookup, and one matrix unit per core (like Arm SME~\cite{sme2022wilkinson}) to offload deep neural network computation. The TMU is an access unit specifically designed to accelerate memory access of \emph{all} sparse/dense tensor algebra expressions. As discussed later in \Cref{sec:dae-abstraction}, by interpreting embedding operations as tensor algebra expressions, we can program the TMU to lookup embedding vectors for the core, a SIMD CPU in this case. In all experiments, we measured performance with full-system gem5~\cite{lowepower2020gem5} and power with McPAT~\cite{li2009mcpat}.

\subsection{Architectural Advantage of DAE Designs} \label{sec:dae-advantage} 

\begin{figure}
\centering
    \includegraphics[width=\linewidth]{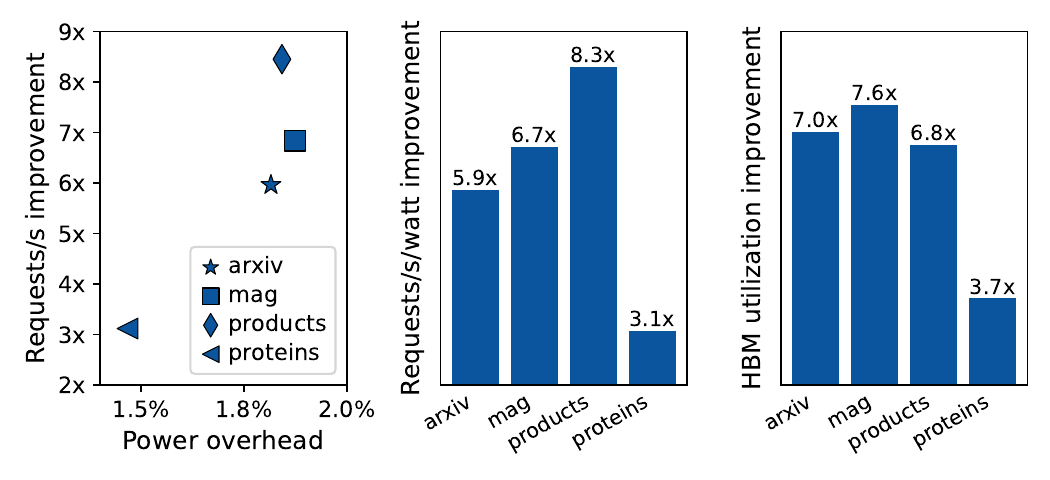}
    \vspace{-1em}
    \begin{subfigure}[b]{0.327\linewidth} 
        \vspace{-4em}
        \caption{}
        \label{fig:ppw-scatter-plot}
    \end{subfigure}
    \begin{subfigure}[b]{0.327\linewidth} 
        \vspace{-4em}
        \caption{}
        \label{fig:ppw-dae-improvement}
    \end{subfigure}
    \begin{subfigure}[b]{0.327\linewidth} 
        \vspace{-4em}
        \caption{}
        \label{fig:hbm-dae-improvement}
    \end{subfigure}
    \caption{Performance and efficiency of a DAE core over a traditional core (details in \Cref{sec:dae-architecture}) on GNN embedding operations (\Cref{sec:graph-learning}). Access units like the TMU can issue and track more memory requests at low power consumption (a), achieving high efficiency (b) and HBM utilization (c).}
    \label{fig:dae-advantage}
\end{figure}

The main advantage of DAE designs is that, by decoupling embedding lookup and computation into two distinct units, these units can be specialized and optimized independently.

For instance, as the TMU implements all the logic to traverse and load tensor operands using specialized dataflow hardware, it can issue memory requests with higher throughput and efficiency than the core~\cite{tmu2023micro}. Moreover, as the TMU can run at a lower frequency, it can track 8$\times$ more outstanding requests, with no timing issues, and low power consumption (less than 2\% overhead)~\cite{tmu2023micro}. Hence, as shown in \Cref{fig:ppw-scatter-plot} and \ref{fig:ppw-dae-improvement}, the TMU achieves 5.7$\times$ and 5.6$\times$ higher requests/s and requests/s/watt than traditional cores, 5.2$\times$ and 6.3$\times$ better than doubling core's outstanding requests. In this way, as shown in \Cref{fig:hbm-dae-improvement}, the TMU utilizes 4$\times$---8$\times$ more memory bandwidth than a traditional core, requiring smaller and fewer cores to saturate the processor's HBM bandwidth.

\begin{table}
  \centering
  \footnotesize
\[
\begin{tabular}{rccc}
\toprule
  \textbf{Property Description} & \textbf{RM1} & \textbf{RM2} & \textbf{RM3} \\
\toprule
  \textbf{Segments per batch per core} & 64 & 32 & 16 \\
  \textbf{Embedding entries per table} & 16K & 16K & 16K \\
  \textbf{Elements per embedding vector} & 32 & 64 & 128 \\
  \textbf{Tables per core} & 2 & 2 & 2 \\
  \textbf{Lookups per segment} & 64 & 128 & 256 \\
\hline
\end{tabular}
\]
  \caption{Tested DLRM models.}
  \label{tab:dlrm-configurations}
  \vspace{-2em}
\end{table}

\begin{figure}
\includegraphics[width=\linewidth]{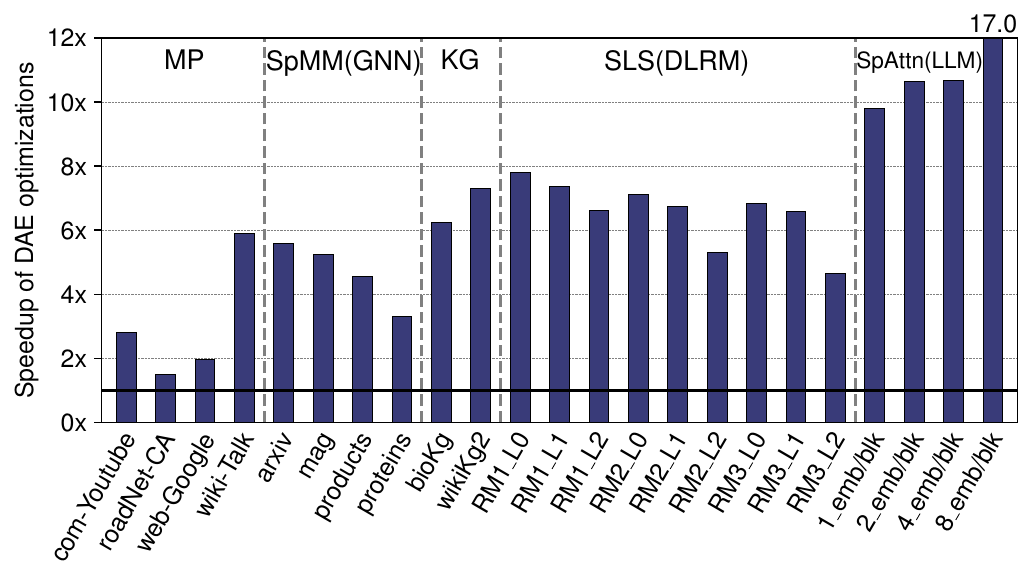} 
\vspace{-2.5em}
\caption{Performance benefit of offloading embedding lookup to a near-core access unit like the TMU (details in \Cref{sec:dae-architecture}). All embedding operations are high-performance multicore implementations from the literature (\Cref{sec:architectural-challenges}). For graph-learning models, the inputs and feature sizes are reported in \Cref{tab:graph-inputs}. For DLRMs, we tested the three configurations in~\Cref{tab:dlrm-configurations}, each one running inputs with low (\texttt{L0}), medium (\texttt{L1}), and high (\texttt{L2}) locality~\cite{dlrm2020hpca}. For LLMs, we tested the original BigBird setting~\cite{bigbird2024neurips} while varying the blocks sizes. Overall, offloading lookup to a specialized access unit improves performance of embedding operations by 5.8$\times$.}
\label{fig:speedups-wrt-cpu}
\end{figure}

From the core's perspective, offloading embedding lookup to the TMU allows to fully utilize the core's resources for compute operations like embedding reductions. In the end, as shown in \Cref{fig:speedups-wrt-cpu}, decoupling embedding lookup from computation on multicore processors achieves an average 5.8$\times$ speedup on the embedding operations in \Cref{sec:characterization-implications}. These speedups are mostly proportional to the temporal locality and compute-per-lookup ratio of the model, all the way to 17$\times$ improvements for SpAttn which has no compute and can be fully offloaded to the TMU.

\subsection{Impact of DAE in Datacenters} \label{sec:gpu-speedups} 

In the remainder of this section, we demonstrate that DAE multicore processors outperform GPUs in end-to-end GNN models. We compare performance against an Nvidia T4 GPU~\cite{nvidiat4}, as it offers similar peak memory bandwidth and computational performance of our DAE multicore processor, and evaluate performance per watt against an Nvidia H100~\cite{nvidiah100}, a widely used solution in datacenters. We focus on GNNs models as (1)~their GPU implementations use high-performance libraries~\cite{ogb2020neurips}, (2)~their input datasets have various localities, and (3)~they pose unique architectural challenges. As GNNs alternate layers of embedding operations and Deep Neural Networks (DNNs), these two operations need to be computed on the same device, as offloading DNNs to a compute accelerator would incur prohibitive host-to-device transfers.

\begin{figure}
    \centering
    \includegraphics[width=\linewidth]{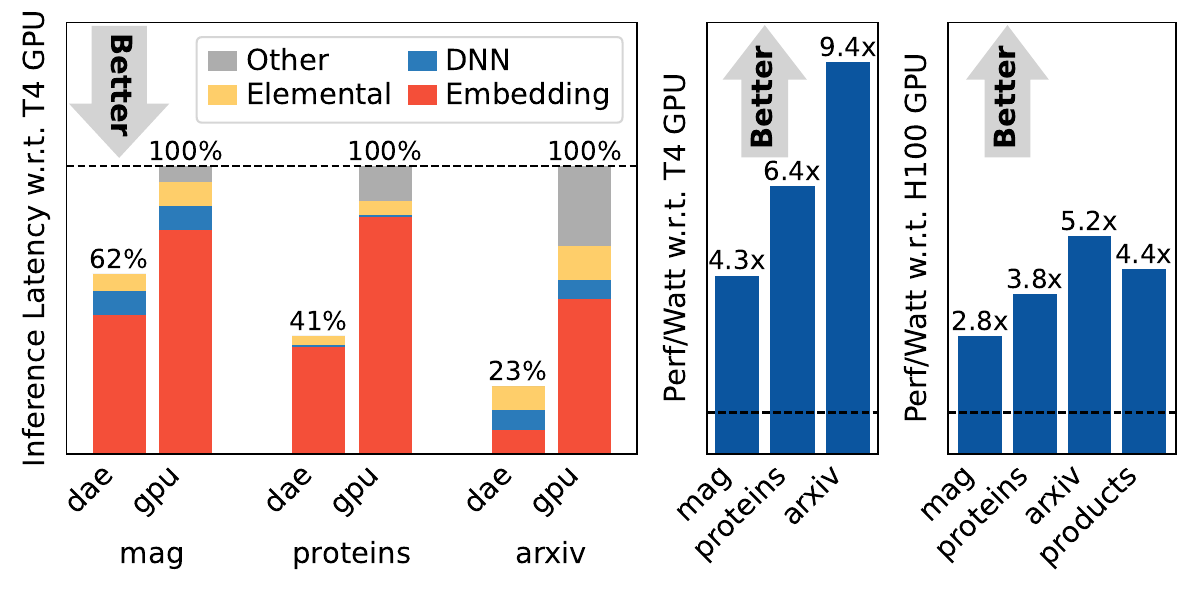}
    \vspace{-1em}
    \begin{subfigure}[b]{0.5\linewidth} 
        \vspace{-3em}
        \caption{}
        \label{fig:t4-lat-comparison}
    \end{subfigure}
    \begin{subfigure}[b]{0.25\linewidth} 
        \vspace{-3em}
        \caption{}
        \label{fig:t4-ppw-comparison}
    \end{subfigure}
    \begin{subfigure}[b]{0.2\linewidth} 
        \vspace{-3em}
        \caption{}
        \label{fig:h100-ppw-comparison}
    \end{subfigure}
    \caption{Performance and efficiency comparison of DAE processors w.r.t. GPUs on OGB GNNs models~\cite{ogb2020neurips} (\texttt{products} did not fit on T4 GPU). For the DAE multicore processor, we measured inference latency with gem5. For the T4 GPU, we used Nvidia Nsys~\cite{nvidiansys}, and only included the kernel execution time and no host-to-device transfer. GPU power consumption is measured with the \texttt{torch.cuda.power\_draw} function whereas the power consumption of the DAE processor is (over)estimated by utilizing the TDP of similar traditional processors~\cite{sapphire2022nassif}. DAE processors outperform GPUs on both inference latency and performance/watt. }
    \label{fig:gpu-comparison}
\end{figure}

\Cref{fig:t4-lat-comparison} shows the inference latency breakdown of GNN models on the DAE processor and the Nvidia T4 GPU~\cite{nvidiat4}. Although the two systems have the same peak memory bandwidth, the DAE system utilizes 4.6$\times$ higher bandwidth, executing embedding operations 1.6$\times$---6.3$\times$ faster than the T4 GPU. As the two systems have similar peak compute, the DNN layers have similar execution time. Hence, the DAE processor achieves 2.6$\times$ higher end-to-end performance than the T4 GPU. As the DAE processor saturates memory bandwidth with only 8 cores, it consumes less power than the T4 GPU, and achieves 6.4$\times$ higher perf/watt, as shown in \Cref{fig:t4-ppw-comparison}. For the same reasons, \Cref{fig:h100-ppw-comparison} demonstrates that the DAE processor also achieves 4$\times$ higher perf/watt than Nvidia's H100 datacenter GPU~\cite{nvidiah100}, which employs a smaller process node than the T4.  

%% file: sec_04_dae_abstraction.tex
\section{DAE Abstraction for Embedding Operations} \label{sec:dae-abstraction}

So far, we have demonstrated the benefits of decoupling embedding lookup from computation. However, such optimization requires a different programming model. In this section, we formalize the DAE programming model with the Decoupled Lookup-Compute (DLC) IR, a low-level programming abstraction for embedding operations on general DAE designs. Similarly to the LLVM IR\cite{llvm2004cs}, the DLC IR provides an abstraction to separate target-agnostic optimizations from target-specific code generation for DAE designs. As discussed in \Cref{sec:ember-overview}, the DLC IR greatly simplifies the Ember design.

\begin{figure}
    \centering
    \includegraphics[width=\linewidth]{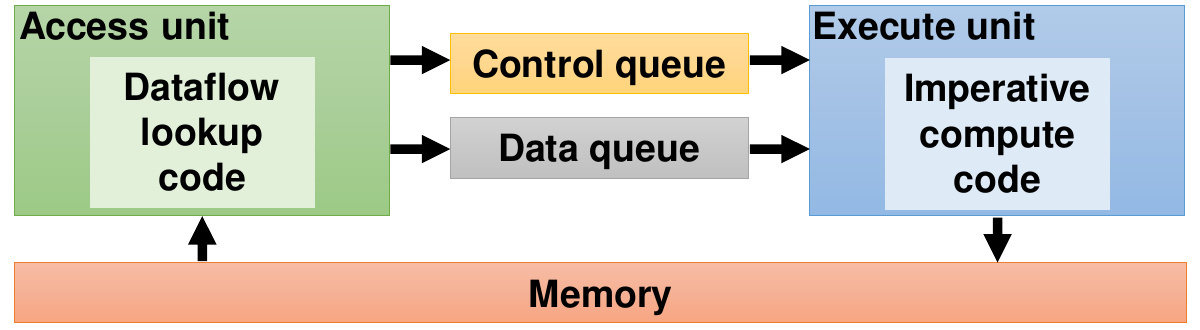}
    \vspace{-2em}
    \caption{DAE architectural abstraction.}
    \label{fig:dae-abstraction}
    \vspace{-0.5em}
\end{figure}

\Cref{fig:dae-abstraction} shows our abstract DAE architecture, which is composed of an access unit to load embedding vectors, an execute unit for computation, and a data queue and control queue to stream data and commands from the access unit to the execute unit. To map embedding operations to such abstraction, we observe that general embedding operations like the ones in \Cref{tab:characterization} are all variants of sparse-dense tensor operations. For instance, the SLS function in \Cref{fig:sls-example} can be interpreted as a sparse-dense matrix multiplication (SpMM) $Z_{i,j}=A_{i,k}B_{k,j}$, with an $ikj$ loop schedule, and sparse matrix stored in a Compressed Sparse-Row (CSR) format~\cite{barret94csr}. For the example in \Cref{fig:irregular-lookup}, dimension $i$ of the sparse matrix $A$ contains the sentences in a batch whereas dimension $k$ contains the possible words in a sentence. $A_{pq}=1$ if the sentence $p$ contains the word $q$, and 0 otherwise. The CSR format stores all the indices of the non-zero elements and the pointers where each sentence begins. Multiplying the sparse matrix $A$ for the dense matrix $B$ accumulates all embedding vectors of the words in each sentence, which is the SLS code in \Cref{fig:sls-imperative}. Non-zero values different than one rescale the embedding vector, which is needed in GNNs (\Cref{sec:graph-learning}). MP models are a Sampled-Dense-Dense Matrix Multiplication (SDDMM) fused with an SpMM kernel~\cite{fusedmm2021ipdps}. KGs are SLS functions that use semirings---general algebraic structures with addition and multiplication---and just have one non-zero per row, not requiring segment pointers or lengths. SpAttn is similar to KG but with a blocked format and no compute.

The Sparse Abstract Machine (SAM)~\cite{sam2023asplos} proves that tensor algebra operations, and hence embedding operations, can be expressed as a dataflow chain of access blocks and compute blocks. Hence, we can map lookup blocks to the access unit and compute blocks to the execute unit, and stream data and commands through the queues~\cite{tmu2023micro}. The DLC IR represents lookup blocks with streaming dataflow code, as it maps well to spatial architectures specialized for irregular memory-intensive workloads~\cite{sam2023asplos,tmu2023micro,spzip2021isca,fpga2022ccpe,topk2021dac,eigen2021fccm,fpga2021iwocl,cad2022tc}. Compute blocks, instead, are represented with imperative code as it can implement all the compute variants introduced above, and others (e.g., quantization and approximate computing). \Cref{fig:sls-abstraction} shows an example of The DLC IR for the SLS function, which is discussed in the reminder of this section.

\input{fig_sls_abstraction}

The DLC lookup code loads embedding elements through memory streams. The iteration space for these memory streams is contained in index streams, which are generated by traversal operators and can be transformed with integer ALU streams. Formally:
\begin{itemize}
    \item \textbf{\texttt{loop\_tr(lb,ub,stride)}}: traverses the iteration space \texttt{i=lb; i<ub;i+=stride}, where \texttt{lb} and \texttt{ub} are streams or immediate values and \texttt{stride} is an immediate value. The stream \texttt{loop\_tr.0} contains the induction variable.
    \item \textbf{\texttt{mem\_str(base,idx)}}: loads into a stream the values of the \texttt{base} memory locations indexed by the \texttt{idx} stream.
    \item \textbf{\texttt{alu\_str(op,op1,op2)}}: computes an integer binary operation \texttt{op} $\in \{ +, -, \times, \div \}$ on operands \texttt{op1} and \texttt{op2}, which can either be streams or immediate values.
\end{itemize}
Lines 1-12 in \Cref{fig:sls-lookup-code} represent  lines 2-6 in \Cref{fig:sls-imperative}.

Compute code consists of callbacks that the access unit triggers while traversing and loading tensor operands. In \Cref{fig:sls-imperative}, we can wrap the FMA operation in line 7 into a callback that the access unit can trigger at each iteration of its parent loop. To trigger such callback, the access unit marshals its corresponding token (the inner-loop iteration token ($e_i$)) in the control queue and its operands (result coordinates $b$ and $e$ and value $v$) in the data queue. Line 14-17 in \Cref{fig:sls-lookup-code} and line 2-4 in \Cref{fig:sls-compute-code} show how these tokens and data are pushed and popped from the queues, whose content is shown in \Cref{fig:sls-queue}. Formally, we can program the access unit to marshal control tokens and operands at each iteration, begin, and end of a fiber traversal through the following operations:
\begin{itemize}
    \item \textbf{\texttt{push\_op(s\_id,tu\_id,event)}}: pushes into the data queue (dataQ) the content of the \texttt{s\_id} stream every time a traversal \texttt{event} $\in \{ \texttt{beg}, \texttt{ite}, \texttt{end} \}$ occurs in \texttt{tu\_id}.
    \item \textbf{\texttt{callback(tu\_id,event)}}: pushes into the control queue (ctrlQ) the corresponding control token every time a traversal \texttt{event} $\in \{ \texttt{beg}, \texttt{ite}, \texttt{end} \}$ occurs in \texttt{tu\_id}.
\end{itemize}
whereas the execute unit reads tokens and operands with:
\begin{itemize}
    \item \textbf{\texttt{pop()}}: pops a control token from the data queue.
    \item \textbf{\texttt{pop<ty>()}}: pops a value with type \texttt{ty} from the data queue.
\end{itemize}
\Cref{fig:opt-impact-idx} shows a more complex example with multiple callbacks. 

In the remainder of this paper, we demonstrate how Ember generates optimized DLC code from PyTorch and TensorFlow embedding operations, enabling full DAE potential at no programmability cost.

%% file: fig_sls_abstraction.tex
\begin{figure}
    \centering

    \begin{subfigure}{\columnwidth}
        \centering
        \includegraphics[width=\columnwidth]{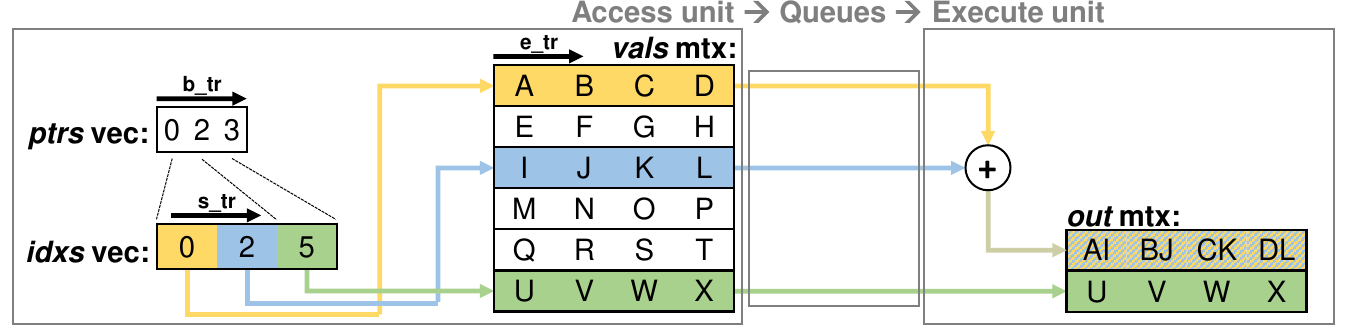}
        \caption{Example of SLS function.}
        \label{fig:sls-example}
    \end{subfigure}
    
    \begin{minipage}{0.5\textwidth}
    \begin{lstlisting}
void sls(idxs, ptrs, vals, out){
  for(int b = 0; b < num_batches; b++){ // Batch traversal (b_tr)
    for(int p = ptrs[b]; p < ptrs[b+1]; p++){ // Segment traversal (s_tr)
      idx = idxs[p]; // Load embedding index
      for(int e = 0; e < emb_len; e++){ // Embedding vector traversal (e_tr)
        val = vals[p,e]; // Load embedding element
        out[b,e] += val; }}}} // Reduce embedding elements
    \end{lstlisting}
    \vspace{-0.5em}
    \subcaption{Imperative code.}
    \label{fig:sls-imperative}
    \end{minipage}

    \begin{minipage}{\columnwidth}
    \begin{lstlisting}
 tu b_tr=loop_tr(0, num_batches, 1) // iterate thru batch
 str beg_ptr=b_tr.mem_str(ptrs, b_tr.ite) // load segment beg ptr
 str end_pos=b_tr.alu_str('+', b_tr.ite, 1) // compute next ptr pos
 str end_ptr=b_tr.mem_str(ptrs, end_pos) // load segment end ptr

 tu s_tr=loop_tr(beg_ptr, end_ptr, 1) // iterate thru segment
 str emb_idx=s_tr.mem_str(s_tr, idxs, s_tr.ite) // load embedding index
 str emb_beg=s_tr.alu_str(s_tr, '*', emb_idx, emb_len) // compute emb addr

 tu e_tr=loop_tr(0, emb_len, 1) // iterate thru emb vec
 str emb_pos= alu_str('+', emb_beg, e_tr.ite) // compute element addr
 str emb_val= mem_str(vals, emb_pos) // load emb element

 push_op(b_tr.ite, e_tr, ite) // push batch position
 push_op(e_tr.ite, e_tr, ite) // push emb el position
 push_op(emb_val, e_tr, ite) // push emb el value
 callback(e_tr, ite) //trigger call on every emb iter
    \end{lstlisting}
    \vspace{-0.5em}
    \subcaption{DLC dataflow lookup code.}
    \label{fig:sls-lookup-code}
    \end{minipage}
    
    \begin{subfigure}{\columnwidth}
        \centering
        \includegraphics[width=0.9\columnwidth]{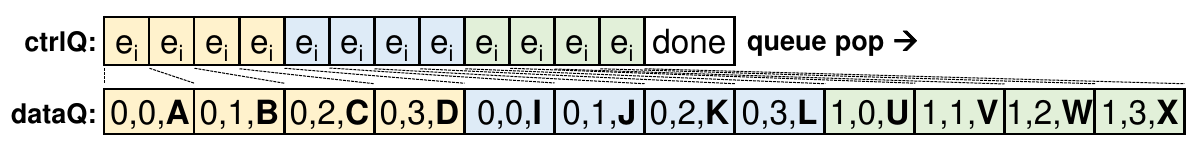}
        \caption{DAE queue content.}
        \label{fig:sls-queue}
    \end{subfigure}
    
    \begin{minipage}{\columnwidth}
    \begin{lstlisting}
while(ctrlQ.pop() != done){ // until some element
  b = dataQ.pop<1 x i32>(); // read embedding position
  e = dataQ.pop<1 x i32>(); // read element position
  v = dataQ.pop<1 x f32>(); // read element value
  fma(&out[b,e],v); // compute and store
    \end{lstlisting}
    \vspace{-0.5em}
    \subcaption{DLC imperative compute code.}
    \label{fig:sls-compute-code}
    \end{minipage}
    \vspace{-1em} 
    \caption{DAE abstraction for the SLS operation.}
    \label{fig:sls-abstraction}
\end{figure}

%% file: sec_05_ember_overview.tex
\section{Ember Overview} \label{sec:ember-overview}

\begin{figure}
    \centering
    \includegraphics[width=0.8\columnwidth]{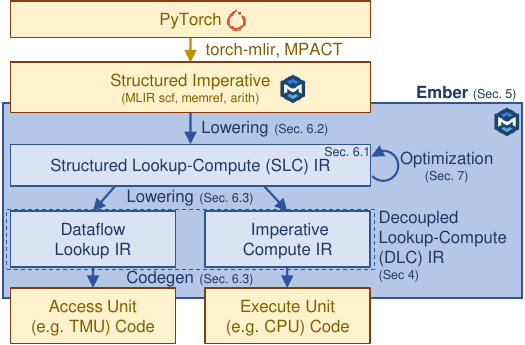}
    \vspace{-1em}
    \caption{Overview of Ember. Yellow represents compiler inputs/outputs and blue represents new contributions.}
    \label{fig:ember-overview}
    \vspace{-0.5em}
\end{figure}

\Cref{fig:ember-overview} shows an MLIR~\cite{mlir2021cgo} implementation of Ember. Overall, Ember compiles PyTorch and TensorFlow embedding operations into optimized DLC code (\Cref{sec:dae-abstraction}), which is then used to generate target-specific access and execute code. However, the DLC IR already separates access code and execute code, which communicate through queue (de)serialization. This breaks the control flow and data flow of the input application, hindering global optimizations such as vectorization or code motion across access and execute code. Ember overcomes such issue with the Structured Lookup-Compute (SLC) IR (\Cref{sec:slc-ir}).

Structured IRs represent for-loops as structured operations, which facilitates analysis and optimizations of complex code like embedding operations. The Structured Control Flow (SCF) IR is perhaps the most popular structured IR in MLIR, and resembles the imperative code in \Cref{fig:sls-imperative}. Ember's SLC IR is the natural extension of the SCF IR for DAE code, and is specifically designed to facilitate global optimizations across access and execute code.

Hence, Ember first generates the SCF IR from PyTorch embedding operations or general tensor algebra expressions with tools like torch-mlir~\cite{torchmlir2024software} or MPACT~\cite{mpact2021google}, respectively. Then, Ember lowers SCF code to the SLC IR and performs global optimizations (\Cref{sec:scf-lowering} and \ref{sec:optimizations}). After lowering to the DLC IR (\Cref{sec:slc-lowering}), access and execute code are further optimized independently, and lowered to target-specific IRs such as the \texttt{tmu} and \texttt{llvm} dialect, which are then used for code generation. All these steps are presented in the following sections.

%% file: sec_06_slc_ir.tex
\section{SCF decoupling and lowering to/from SLC IR} \label{sec:decoupling}

\input{fig_slc_ir}

\input{fig_scf_slc_lowering}

This section introduces the SLC IR, and how Ember lowers SCF and DLC code to and from it, respectively.

\subsection{The SLC IR} \label{sec:slc-ir}

\Cref{fig:slc-ir} shows the SLC grammar whereas \Cref{fig:lowering-slc} shows the SLC IR of the SLS function. The SLC IR has a similar structure to SCF code, besides (1)~SCF loops, index arithmetic, and load operations to offload to the access unit are represented as SLC loops and streams and (2)~compute code for the execute unit is wrapped into SLC callbacks within the SLC loops. In this way, callbacks can access data from streams with stream-to-value conversion operations, preserving the data flow of the embedding operation which is no longer (de)serialized through queues. This enables Ember to simultaneously optimize lookup and compute code through global analyses and transformations, including dominance analysis, vectorization, and code motion across different compute regions as well as across access and execute code, as discussed in \Cref{sec:optimizations}.

\subsection{Lowering the SCF IR to the SLC IR} \label{sec:scf-lowering}

\Cref{fig:lowering} shows a lowering example from SCF to SLC. To generate the SLC IR, Ember recursively traverses the loop hierarchy of the SCF code looking for loops to offload to the access unit. We define an \emph{offloading candidate} as a loop that (1)~has iteration bounds which are either static or computed by another offloading candidate and (2)~loads from at least one read-only memory location which has not been read from a parent loop. Condition (1) is necessary as access units cannot generally read data from the execute units. Condition (2) excludes workspace loops~\cite{workspaces2019cgo,spworkspaces2024pldi} (i.e. loops that just work on partial results). As partial results are likely cached, workspace loops do not benefit from memory acceleration. All loops in \Cref{fig:lowering-scf} are offloading candidates. Instead, the last two MP lines in \Cref{tab:characterization} are workspaces loops, as they just multiply and accumulate the temporary vector with the vertex vector, which has been read already.

As embedding operations are variants of sparse-dense tensor multiplications (\Cref{sec:dae-abstraction}), their loop hierarchy can only have at most one offloading candidate per level and at most one workspace loop per level~\cite{taco2017oopsla,workspaces2019cgo}. Hence, the decoupling algorithm recursively traverses and selects one offloading candidate per level, leaving the other loops for software execution. The SCF offloading candidates are lowered to SLC for-loops.

Then, Ember places callbacks in the SLC loops. Operations to be offloaded (i.e., read-only load operations and index arithmetic) are transformed into streams and moved before their corresponding callback. Operations not to be offloaded (compute code and workspace loops) are moved inside their corresponding callback, which is executed in software. For instance, the read-only load operation in line 11 \Cref{fig:lowering-scf} is transformed and moved before the callback whereas the accumulation operations in line 12 and 13 are moved inside the callback. Finally, stream-to-value operations are added for each operation in a callback reading from a stream.

\subsection{Lowering SLC IR to DLC IR} \label{sec:slc-lowering}

After applying optimizations (\Cref{sec:optimizations}), the SLC IR is lowered to the low-level DLC IR (\Cref{sec:dae-abstraction}). The SLC IR is generated by traversing the SLC IR from the outer to the inner SLC for-loop. SLC for-loops and streams are lowered to DLC traversal operators and streams. Callbacks, instead, are moved into the while-loop in the compute code (\Cref{fig:dae-abstraction}). Multiple callbacks are chained into an if-then-else construct that cases the token IDs popped from the control queue (e.g. \Cref{fig:opt-impact-idx}). Token's push instructions are generated according to the location of the callback in the SLC IR. Operands' push and pop instructions are generated according to the SLC stream-to-value operations.

%% file: fig_slc_ir.tex
\begin{figure}
  \footnotesize
\[
\begin{array}{rl|@{\hspace{1.5\arraycolsep}}rl}
\textit{Any CPU Statement} & STMT &
\textit{Memory Reference} & mref \\
\textit{CPU variable} & var &
\textit{Stream Variable} & s \\
\textit{Unsigned Integer} & \text{uint} \\
\end{array}
\]
\[
\begin{array}{rlrl}
  \textit{For-Loop} & FOR & \bnfdef & \text{for } (HEAD) \text{ \{ } SDEC^* ~ BODY \text{ \} }\\
  \textit{For Header} & HEAD & \bnfdef & s = r \text{ to } r \text{ step } \text{uint} \\
  \textit{Stream Declaration} & SDEC  & \bnfdef & s = se \\
  \textit{Loop Body} & BODY & \bnfdef & CALL ~\bnfalt~ CALL? ~~ FOR ~~ CALL? \\
  \textit{Callback} & CALL & \bnfdef & \text{callback()} \text{ \{ } TVAL^*~ STMT^+ \text{ \}} \\
  \textit{Value Conversion} & TVAL & \bnfdef & var = \text{to\_val}(s) \\
  \textit{Loop Range} & r & \bnfdef &  s ~\bnfalt~ var ~\bnfalt~ \text{uint}  \\
  \textit{Stream Expression} & se & \bnfdef & ls~\bnfalt~as~\bnfalt~cs~\bnfalt~bs \\
  \textit{Load Stream} & ls  & \bnfdef & \text{load\_str} (mref[indices^*], uint) \\
  \textit{Indices} & indices & \bnfdef & s~\bnfalt~var~\bnfalt~ \text{uint} \\
  \textit{ALU Stream} & as & \bnfdef & \text{alu\_str} (sop, s_1, s_2) \\
  \textit{Stream Ops} & sop  & \bnfdef & +~\bnfalt~-~\bnfalt~*~\bnfalt~ 
/~\bnfalt~\%~\bnfalt~ \ldots  \\
\end{array}
\]
  \vspace{-1em}
  \caption{The Structured Lookup-Compute (SLC) IR.}
  \label{fig:slc-ir}
  \vspace{-0.5em}
\end{figure}

%% file: fig_scf_slc_lowering.tex
\begin{figure}
    \noindent
    \begin{subfigure}[]{0.93\columnwidth}
        \begin{subfigure}[t]{0.47\columnwidth} 
            \begin{lstlisting}[xleftmargin=0em]
void sls(idxs: mref<? x idx>,
         ptrs: mref<? x idx>,
         vals: mref<? x f32>,
         out: mref<? x ? x f32>){
 (*\hlloop{for(idx b=0; b<n\_batches; b++)}*){
  (*\hlmem{idx beg=ptrs[b]}*);
  (*\hlmem{idx end=ptrs[b+1]}*);
  (*\hlloop{for(idx p=beg; p<end; p++)}*){
   (*\hlmem{idx i=idxs[ptr]}*);
   (*\hlloop{for(idx e=0; e<emb\_len; e++)}*){
    (*\hlmem{f32 val=vals[i,e]}*);
    (*\hlcompute{f32 acc=out[b,e]}*);
    (*\hlcompute{out[b,e]=acc+val}*); }}}}
            \end{lstlisting}
            \vspace{2.9em}
            \caption{SCF IR}
            \label{fig:lowering-scf}
        \end{subfigure}
        \begin{subfigure}[t]{0.55\columnwidth} 
            \begin{lstlisting}[xleftmargin=0em]
void sls(idxs: mref<? x idx>,
         ptrs: mref<? x idx>,
         vals: mref<? x f32>,
         out: mref<? x ? x f32>){
 (*\hlloop{slc.for(str s\_b from 0 to n\_batches)}*){
  (*\hlmem{str s\_beg=slc.mem\_str(ptrs[s\_b])}*);
  (*\hlmem{str s\_end=slc.mem\_str(ptrs[s\_b+1])}*);
  (*\hlloop{slc.for(str s\_p from s\_beg to s\_end)}*){
   (*\hlmem{str s\_i=slc.mem\_str(idxs[s\_p])}*);
   (*\hlloop{slc.for(str s\_e from 0 to emb\_len)}*){
    (*\hlmem{str s\_val=slc.mem\_str(vals[s\_i,s\_e])}*);
    slc.callback{
     idx b=slc.to_val(s_b);
     idx e=slc.to_val(s_e);
     f32 val=slc.to_val(s_val);
     (*\hlcompute{f32 acc=out[b,e]}*);
     (*\hlcompute{out[b,e]=acc+val}*); }}}}}
            \end{lstlisting}
            \caption{SLC IR}
            \label{fig:lowering-slc}
        \end{subfigure}
    \end{subfigure}
    \vspace{-1em}
    \caption{Lowering step from SCF IR to SLC IR. Ember converts SCF \hlloop{for-loops} into SLC for-loops and SCF \hlmem{memory accesses} (loads and index computation) into SLC streams. Ember wraps \hlcompute{computation} into SLC callbacks that can access SLC streams with stream-to-value operations.}
    \label{fig:lowering}
\end{figure}

%% file: sec_07_optimizations.tex
\section{Optimizing Embedding Operations} \label{sec:optimizations}

\input{fig_opt_impact}

\input{fig_opt_transformations}

This section describes three key DAE optimizations enabled by the SLC IR. \Cref{fig:opt-transformations} and \ref{fig:opt-impact} show how these optimizations transform the SLC IR, compute code, and queues.

\subsection{Vectorization} \label{sec:vectorization}

Vectorization is one of the most impactful DAE optimizations~\cite{tmu2023micro}. As shown in \Cref{fig:opt-impact-vec}, for each token, vectorization pops a vector of \textit{vector length (vlen)} elements, improving marshaling and compute efficiency. While auto-vectorizing general code is challenging~\cite{comp2001pub}, the SLC IR provides an effective representation to vectorize loading, marshaling, and computing of embedding operations.

As shown in \Cref{fig:opt-vec}, Ember vectorizes code by converting SLC operations into their vectorized SLCV duals. Conversely from the SLC for-loop, the SLCV for-loop (1) adds a \emph{vector length} attribute, (2) instantiates vectorized induction variables, and (3) introduces the concept of \texttt{mask} to handle loop boundaries not multiple of the vector length. Masks are used by SLCV streams to perform vectorized index computation and data loading.

We define a \emph{vectorization scheme} as the set of for-loops within a parent loop $p$ and the inner loop $i$, with $p \geq i$, that we intend to vectorize. A for-loop can be vectorized if and only if all of its callbacks can be vectorized. A vectorization scheme is \emph{legal} if and only if all of its for-loops can be vectorized.

Vector extensions such as Arm SVE~\cite{sve2017micro} provide instructions to vectorize most callbacks in embedding operations, making the space of legal vectorization schemes large. However, prior work has demonstrated that the most efficient vectorization scheme for sparse-dense tensor multiplication is inner-loop vectorization, assuming the dense tensor is in row-major order and has rows which are larger than the vector length~\cite{tmu2023micro}. Embedding operations generally satisfy these assumptions (\Cref{sec:architectural-challenges}). Hence, similarly to the MLIR sparsifier~\cite{mlirsparse2022taco}, Ember only attempts inner-loop vectorization.

If the inner for-loop is legal, Ember vectorizes its access and execute code in two steps. Ember starts by vectorizing the inner for-loop and its streams. As the stream-to-value operations in the callbacks expect stream types, the algorithm adds a temporary cast operation for source materialization. Then, during the callback vectorization step, Ember recursively vectorizes the uses of these cast operations, producing full SLCV code. To keep the conversion step simple, load and store operations into callbacks are firstly converted to vector gather and scatter operations, with vector indices and masks coming from its parent vectorized for-loop operation. A further transformation pass simplifies these operations into contiguous vector loads and stores. 

\subsection{Bufferization} \label{sec:bufferization}

Bufferization allows to marshal and compute embedding vectors as compound types. As shown in \Cref{fig:opt-impact-buf}, the access unit pushes, in the control queue, one $e_e$ (embedding-vector end) token for each embedding vector and, in the data queue, the position of the output embedding vector and all of its values. As the length of the embedding vector (\texttt{emb\_len}) is constant, once the core reads an $e_e$ token, it pops \texttt{emb\_len} elements with a vectorized for-loop. Bufferization greatly improves marshaling and compute efficiency, especially for long embedding vectors.

As shown in \Cref{fig:opt-buf}, Ember bufferizes code by firstly initializing a buffer stream of vector type (line 10) before the inner SLCV for-loop, where such loop can push the loaded embedding elements (line 14). Then, Ember moves the inner-loop body callback (line 14-19 in \Cref{fig:opt-vec}) right after the loop (line 16-22 in \Cref{fig:opt-buf}), adds a stream-to-value operation for the buffer (line 18), and adds a loop to iterate through it (line 19-22).

\subsection{Queue alignment} \label{sec:queue-align}

Vector loads are more efficient when aligned to cache lines. However, as shown in \Cref{fig:opt-impact-buf}, scalar operands like segment IDs hinder alignment of embedding vectors. Ember tries to align embedding accesses with different optimizations depending on the code characteristics.

For simpler functions like the SLS in \Cref{fig:opt-buf} where all segment IDs are just loop induction variables, Ember stores a reference of these indices in the core and increments them after the loop. This increment is triggered by a segment-end token, $s_e$ in \Cref{fig:opt-impact-idx}.

As shown in \Cref{fig:opt-idx}, Ember implement such optimization by looking into iteration callbacks for stream-to-value operations that just read the induction variable of their own loop (e.g. line 17 in \Cref{fig:opt-buf}). Then Ember adds a new variable in the SLC loop (\texttt{i}~variable in line 12 in \Cref{fig:opt-idx}), replaces all uses of the stream-to-value operations with that variable (line 20-21), and increments it in the end callback of its child loop (line 22).

However, for more complex models like MP, certain scalars cannot be simplified (e.g. rescaling values). In this case, Ember preserves alignment by padding scalars to vectors while generating the DLC IR. As a further optimization, instead of sending segment IDs, Ember offloads to the access unit full index calculation of partial and output results and directly sends addresses to the core, reducing pressure on core's ALUs. 

\subsection{Model-Specific Optimizations} \label{sec:other-optimizations}

While the optimization we presented so far are for general embedding operations, Ember can also implement model-specific optimizations. For instance, block-sparse attention mechanisms (\Cref{sec:llms}) exhibit (1)~large structured reuse within each block, (2)~low reuse throughout blocks, and (3)~no computation. Hence, we can add \emph{store} streams to write directly into memory without passing through the core. Moreover, we can extend load streams to (1)~select what cache level to read from and (2)~whether to issue temporal or non-temporal requests. As demonstrated in \Cref{sec:model-specific-optimizations-impact}, this greatly improves core and cache utilization. Other optimizations include, for instance, introducing \emph{toggle callback}s to trigger computation on sparse coordinate formats~\cite{mttkrp2019jsc} or \emph{accumulation stream}s to track boundaries of SLS segments by accumulating lengths, instead of using offsets~\cite{dheevatsa2022swhwdlrm}.

%% file: fig_opt_impact.tex
\begin{figure}
    \begin{minipage}{\columnwidth}
        \begin{minipage}{0.5\columnwidth}
        \vfill
        \centering
        \begin{lstlisting}
while(ctrlQ.pop() != done){
  b = dataQ.pop<1 x index>();
  e = dataQ.pop<1 x index>();
  v = dataQ.pop<1 x f32>();
  fma(&out[b,e],v); }
        \end{lstlisting}
        \vfill
        \end{minipage}
        \begin{minipage}{0.5\columnwidth}
        \vfill
        \centering
        \includegraphics[width=\linewidth, trim={0 3in 0 0},clip]{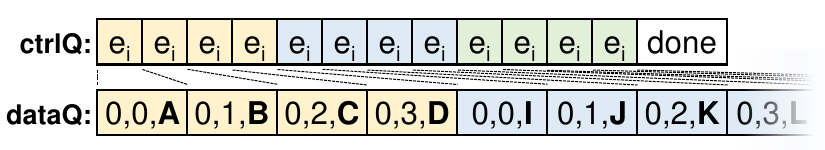}
        \vfill
        \end{minipage}
    \vspace{-0.5em}
    \subcaption{Unoptimized code.}
    \label{fig:opt-impact-none}
    \end{minipage}
    
    \begin{minipage}{\columnwidth}
        \begin{minipage}{0.5\textwidth}
        \vfill
        \centering
        \begin{lstlisting}
while(ctrlQ.pop() != done){
  b = dataQ.pop<1 x index>();
  e = dataQ.pop<1 x index>();
  v = dataQ.pop<vlen x f32>();
  v_fma(&out[b,e], v);
        \end{lstlisting}
        \vfill
        \end{minipage}
        \begin{minipage}{0.5\textwidth}
        \vfill
        \centering
        \includegraphics[width=\linewidth, trim={0 2in 0 1in},clip]{img_queue_optimizations_impact.pdf}
        \vfill
        \end{minipage}
    \vspace{-0.5em}
    \subcaption{Vectorized code.}
    \label{fig:opt-impact-vec}
    \end{minipage}

    \begin{minipage}{\columnwidth}
        \begin{minipage}{0.5\textwidth}
        \vfill
        \centering
        \begin{lstlisting}
while(ctrlQ.pop() != done){
  b = dataQ.pop<1 x index>()
  for(e=0; e<emb_len; e+=vlen){
    v = dataQ.pop<vlen x f32>()
    v_fma(&out[b,e],v) }
        \end{lstlisting}
        \vfill
        \end{minipage}
        \begin{minipage}{0.5\textwidth}
        \vfill
        \centering
        \includegraphics[width=\linewidth, trim={0 1in 0 2in},clip]{img_queue_optimizations_impact.pdf}
        \vfill
        \end{minipage}
    \vspace{-0.5em}
    \subcaption{Bufferized code.}
    \label{fig:opt-impact-buf}
    \end{minipage}

    \begin{minipage}{\columnwidth}
        \begin{minipage}{0.5\textwidth}
        \vfill
        \centering
        \begin{lstlisting}
out_ptr = out
while(tkn = ctrlQ.pop() != done){
  if(tkn == (*$e_e$*)){
    for(e=0; e<emb_len; e+=vlen){
      v = dataQ.pop<vlen x f32>()
      v_fma(out_ptr+e, v) }}
  else if(tkn == (*$s_e$*)){
    out_ptr += emb_len }
        \end{lstlisting}
        \vfill
        \end{minipage}
        \begin{minipage}{0.5\textwidth}
        \vfill
        \centering
        \includegraphics[width=\linewidth, trim={0 0 0 3in},clip]{img_queue_optimizations_impact.pdf}
        \vfill
        \end{minipage}
    \vspace{-0.5em}
    \subcaption{Queue aligned code.}
    \label{fig:opt-impact-idx}
    \end{minipage}
    
    \vspace{-1em} 
    \caption{Impact of SLC optimizations on the SLS compute code and output queues.}
    \label{fig:opt-impact}
\end{figure}

%% file: fig_opt_transformations.tex
\begin{figure}
    \centering
    \begin{minipage}{0.5\textwidth}
    \begin{lstlisting}
void sls(idxs: mref<? x index>, offs: mref<? x index>,
         vals: mref<? x f32>, out: mref<? x ? x f32>){
  //Access: Iterate over segments in a batch
  slc.for(stream s_b from 0 to num_batches){
    stream s_beg = slc.mem_str(offs[s_b]);
    stream s_end = slc.mem_str(offs[s_b+1]);
    //Access: Iterate over embeddings in a segment
    slc.for(stream s_ptr from s_beg to s_end){
      stream s_idx = slc.mem_str(idxs[s_ptr]);
      //Access: Iterate over embedding vector elements
      slc.for(stream s_e from 0 to emb_len){
        stream s_val = slc.mem_str(vals[s_idx,s_e]);
        //Execute: Reduce embedding vectors
        slc.callback{
          index b = slc.to_val(s_b);
          index e = slc.to_val(s_e);
          f32 val = slc.to_val(s_val);
          f32 acc = out[b,e];
          out[b,e] = acc + val; }}}}}
    \end{lstlisting}
    \vspace{-0.7em}
    \subcaption{Unoptimized code.}
    \label{fig:opt-none}
    \end{minipage}

    \begin{minipage}{0.5\textwidth}
    \begin{lstlisting}
(*\unchanged{void sls(idxs: mref<? x index>, offs: mref<? x index>, }*)
         (*\unchanged{vals: mref<? x f32>, out: mref<? x ? x f32>)\{ }*)
  (*\unchanged{//Access: Iterate over segments in a batch }*)
  (*\unchanged{slc.for(stream s\_b from 0 to num\_batches)\{ }*)
    (*\unchanged{stream s\_beg = slc.mem\_str(offs[s\_b]); }*)
    (*\unchanged{stream s\_end = slc.mem\_str(offs[s\_b+1]); }*)
    (*\unchanged{//Access: Iterate over embeddings in a segment }*)
    (*\unchanged{slc.for(stream s\_ptr from s\_beg to s\_end)\{ }*)
      (*\unchanged{stream s\_idx = slc.mem\_str(idxs[s\_ptr]); }*)
      (*\unchanged{//Access: Iterate over embedding vector elements }*)
      slcv.for<vlen>((stream s_e, stream msk) from 0 to emb_len){
        stream s_val = slcv.mem_str<vlen>(vals[s_idx,s_e], msk);
        //Execute: Reduce embedding vectors
        slcv.callback{
          (*\unchanged{index b = slc.to\_val(s\_b); }*)
          index e = slcv.to_val(s_e)[0];
          vec<vlen x f32> val = slcv.to_val<vlen>(s_val);
          vec<vlen x f32> acc = vload<vlen>(out[b,e]);
          vstore<vlen>(acc + val, out[b,e], acc); }}}}}
    \end{lstlisting}
    \vspace{-0.7em}
    \subcaption{Vectorized code.}
    \label{fig:opt-vec}
    \end{minipage}

    \begin{minipage}{0.5\textwidth}
    \begin{lstlisting}
(*\unchanged{void sls(idxs: mref<? x index>, offs: mref<? x index>, }*)
         (*\unchanged{vals: mref<? x f32>, out: mref<? x ? x f32>)\{ }*)
  (*\unchanged{//Access: Iterate over segments in a batch }*)
  (*\unchanged{slc.for(stream s\_b from 0 to num\_batches)\{ }*)
    (*\unchanged{stream s\_beg = slc.mem\_str(offs[s\_b]); }*)
    (*\unchanged{stream s\_end = slc.mem\_str(offs[s\_b+1]); }*)
    (*\unchanged{//Access: Iterate over embeddings in a segment }*)
    (*\unchanged{slc.for(stream s\_ptr from s\_beg to s\_end)\{ }*)
      (*\unchanged{stream s\_idx = slc.mem\_str(idxs[s\_ptr]); }*)
      stream<vec<vlen x f32>> buf = slcv.buf_str(); // Buffer stream
      (*\unchanged{//Access: Iterate over embedding vector elements }*)
      (*\unchanged{slcv.for<vlen>((stream s\_e, stream msk) from 0 to emb\_len)\{ }*)
        (*\unchanged{stream s\_val = slcv.mem\_str<vlen>(vals[s\_idx,s\_e], msk); }*)
        slc.push(buf, s_val); } // Push into the buffer
      (*\unchanged{//Execute: Reduce embedding vectors }*)
      slcv.callback{ // Callback moved at the end of inner loop
        (*\unchanged{index b = slc.to\_val(s\_b); }*)
        vec<vlen x f32> buf_vec = slc.to_val(buf); // Get buffer
        for(index e = 0; e < emb_len; e++){ // Iterate buffer
          vec<vlen x f32> val = buf_vec[e]; // Get buffered element
          (*\unchanged{vec<vlen x f32> acc = vload<vlen>(out[b,e]); }*)
          (*\unchanged{vstore<vlen>(acc + val, out[b,e], acc); \}\}\}\}\}\} }*)
    \end{lstlisting}
    \vspace{-0.7em}
    \subcaption{Bufferized code.}
    \label{fig:opt-buf}
    \end{minipage}

    \begin{minipage}{0.5\textwidth}
    \begin{lstlisting}
(*\unchanged{void sls(idxs: mref<? x index>, offs: mref<? x index>, }*)
         (*\unchanged{vals: mref<? x f32>, out: mref<? x ? x f32>)\{ }*)
  (*\unchanged{//Access: Iterate over segments in a batch }*)
  (*\unchanged{slc.for(stream s\_b from 0 to num\_batches)\{ }*)
    (*\unchanged{stream s\_beg = slc.mem\_str(offs[s\_b]); }*)
    (*\unchanged{stream s\_end = slc.mem\_str(offs[s\_b+1]); }*)
    (*\unchanged{//Access: Iterate over embeddings in a segment }*)
    (*\unchanged{slc.for(stream s\_ptr from s\_beg to s\_end)\{ }*)
      (*\unchanged{stream s\_idx = slc.mem\_str(idxs[s\_ptr]); }*)
      (*\unchanged{stream<vec<vlen x f32>> buf = slcv.buf\_str(); }*)
      (*\unchanged{//Access: Iterate over embedding vector elements }*)
      slcv.for<vlen>((stream s_e, stream msk) from 0 to emb_len)(index i=0) {
        (*\unchanged{stream s\_val = slcv.mem\_str<vlen>(vals[s\_idx,s\_e], msk); }*)
        (*\unchanged{slc.push(buf, s\_val); \} }*)
      (*\unchanged{//Execute: Reduce embedding vectors }*)
      (*\unchanged{slcv.callback\{ }*)
        (*\unchanged{vec<vlen x f32> buf\_vec = slc.to\_val(buf); }*)
        (*\unchanged{for(index e = 0; e < emb\_len; e++)\{ }*)
          (*\unchanged{vec<vlen x f32> val = buf\_vec[e]; }*)
          vec<vlen x f32> acc = vload<vlen>(out[i,e]);
          vstore<vlen>(acc + val, out[i,e], acc); }}}}
    slc.callback{ i++; }}} // end of indices in segment loop
    \end{lstlisting}
    \vspace{-0.7em}
    \subcaption{Queue aligned code.}
    \label{fig:opt-idx}
    \end{minipage}
    
    \vspace{-1em} 
    \caption{Progressive optimization of the SLS SLC IR.}
    \label{fig:opt-transformations}
\end{figure}

%% file: sec_08_evaluation.tex
\section{Evaluation} \label{sec:evaluation}

\definecolor{Gray}{gray}{0.85}
\begin{table}
  \centering
  \footnotesize
\[
\begin{tabular}{lll}
\toprule
  \textbf{Name} & \textbf{MLIR dialects} & \textbf{Description} \\
\toprule
  \texttt{\textbf{emb-opt0}} & \texttt{slc}, \texttt{scf}, \texttt{memref}, \texttt{arith} & unoptimized Ember DAE code \\
  \texttt{\textbf{emb-opt1}} & \texttt{slcv}, \texttt{scf}, \texttt{memref}, \texttt{arith}, \texttt{vector} & \texttt{emb-opt0}+vectorization \\
  \texttt{\textbf{emb-opt2}} & \texttt{slcv}, \texttt{scf}, \texttt{memref}, \texttt{arith}, \texttt{vector} & \texttt{emb-opt1}+bufferization \\
  \texttt{\textbf{emb-opt3}} & \texttt{slcv}, \texttt{scf}, \texttt{memref}, \texttt{arith}, \texttt{vector} & \texttt{emb-opt2}+queue alignment \\
  \texttt{\textbf{ref-dae}}  & \texttt{tmu}, \texttt{scf}, \texttt{memref}, \texttt{arith}, \texttt{vector} & hand-optimized TMU-CPU code \\
\bottomrule
\end{tabular}
\]
  \caption{Evaluated code and reference. Ember's input is torch-mlir~\cite{torchmlir2024software}, output is TMU-CPU machine code.}
  \label{tab:software-versions}
  \vspace{-2em}
\end{table}

In this section, we demonstrate how Ember enables full DAE potential at scale. Our evaluation firstly demonstrates the benefit of the SLC IR by performing an ablation study on the general embedding optimizations (vectorization, bufferization, and queue alignment) introduced in \Cref{sec:optimizations}. Then, it demonstrates the generality of the SLC IR by showing the impact of the model-specific optimizations for LLMs introduced in \Cref{sec:other-optimizations}. Finally, it demonstrates that Ember optimizations match the performance of handwritten code specifically optimized for our TMU-CPU target system.

The experimental setting in this section follows the DAE study in \Cref{sec:dae-advantage}. \Cref{tab:software-versions} summarizes the code utilized in the experiments.

\subsection{Impact of General Optimizations} \label{sec:ablation} 

\begin{figure}
\includegraphics[width=\linewidth]{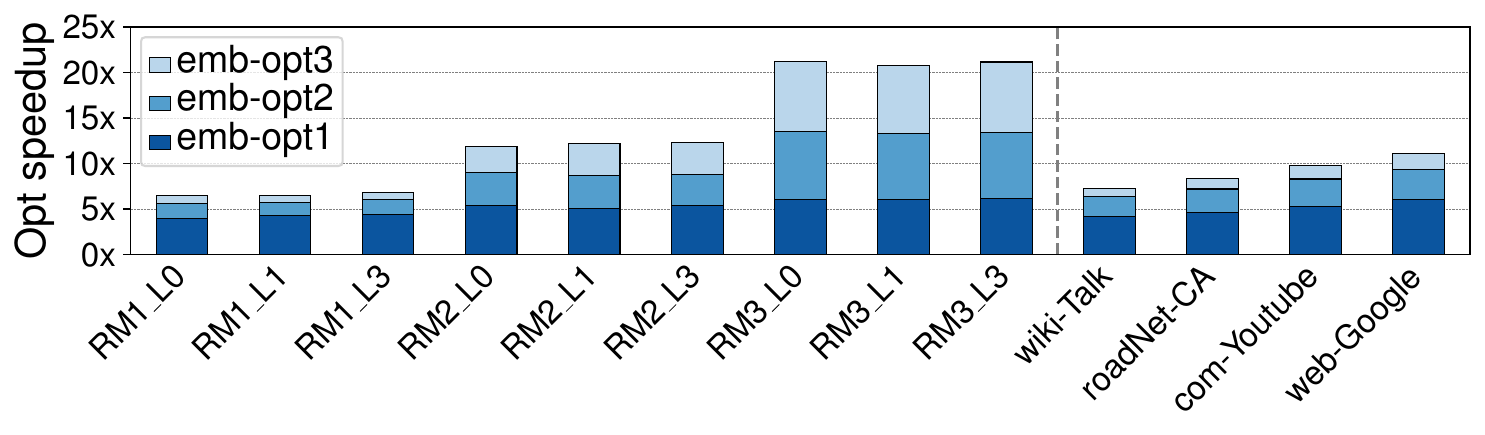}
\vspace{-2em}
\caption{Performance speedup of Ember optimizations on MP models and SLS function for various DLRM models (\Cref{tab:dlrm-configurations}) and input locality. All optimizations combined (\texttt{emb-opt3}) improve performance by 6.6$\times$--21$\times$.}
\label{fig:ablation}
\end{figure}

\begin{figure}
\includegraphics[width=\linewidth]{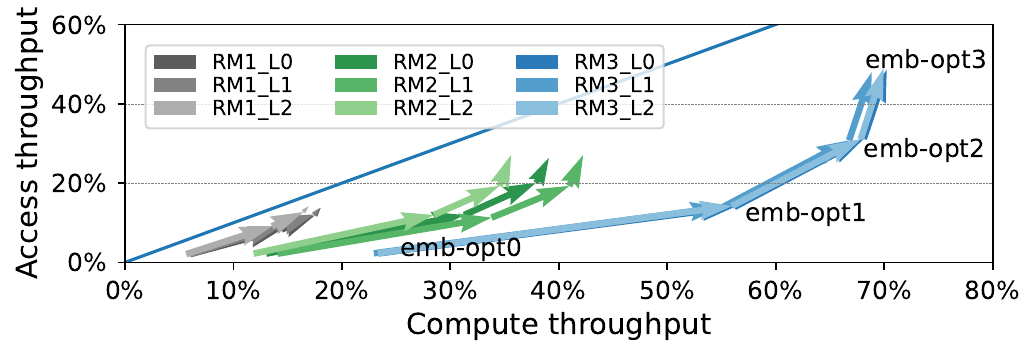}
\vspace{-2em}
\caption{Impact of Ember optimizations on access throughput (TMU) and compute throughput (CPU). By optimizing both, Ember achieves highest performance (top-right). }
\label{fig:outq-throughput-improvement}
\end{figure}

\Cref{fig:ablation} shows the performance impact of general embedding optimizations such as vectorization (\texttt{emb-opt1}), bufferization (\texttt{emb-opt2}), and queue alignment (\texttt{emb-opt3}) over unoptimized Ember-generated code (\texttt{emb-opt0}) for the SLS function and more compute-intensive MP models.

For the SLS function, we evaluated the three DLRMs in~\Cref{tab:dlrm-configurations}, each one running representative synthetic inputs~\cite{dlrm2020hpca} with low (\texttt{L0}), medium (\texttt{L1}), and high (\texttt{L2}) locality. Overall, all optimizations combined (\texttt{emb-opt3}) achieve 6.6$\times$, 12.1$\times$, and 21$\times$ better performance over unoptimized code (\texttt{emb-opt0}) for \texttt{RM1}, \texttt{RM2}, and \texttt{RM3}, respectively. Vectorization is consistently the most impactful optimization with a 5.13$\times$ speedup and only 17\% deviation, whereas other optimizations deliver widely-different performance improvements on different configurations.

To better understand these results, \Cref{fig:outq-throughput-improvement} shows how these optimizations impact the throughput at which the compute unit reads and the access unit writes into the L2 queue. Compute optimizations move upward whereas memory optimizations move rightward. The blue line indicates where the compute-unit throughput equals the access-unit throughput. Only optimizing compute code cannot move above the blue line as the access unit would not be able to marshal enough data to process. This would only improve throughput up to 8$\times$ (\texttt{RM3}, \texttt{emb-opt0}) before the access unit starts to be the bottleneck (blue line). However, because of the SLC IR, Ember can perform global optimizations on both access and compute code and move both rightward and upward in the plot, improving performance by up to 21$\times$ (\texttt{RM3}, \texttt{emb-opt3}). For \texttt{RM1}, the most control-intensive model (shorter loops), vectorization already saturates throughput, leaving other optimizations little room for improvement. For \texttt{RM2} and \texttt{RM3}, by reducing coordinate overhead, bufferization helps to move closer to the blue line. For \texttt{RM3}, the model with largest loops, queue alignment removes index overhead, pushing performance closer to the limit.

As shown in \Cref{fig:ablation}, MP models have similar trends, and the optimization impact is mostly proportional to their compute-per-lookup ratio (\Cref{tab:characterization}).

\subsection{Impact of Model-Specific Optimizations} \label{sec:model-specific-optimizations-impact} 

\begin{figure}
    \centering
    \includegraphics[width=\linewidth]{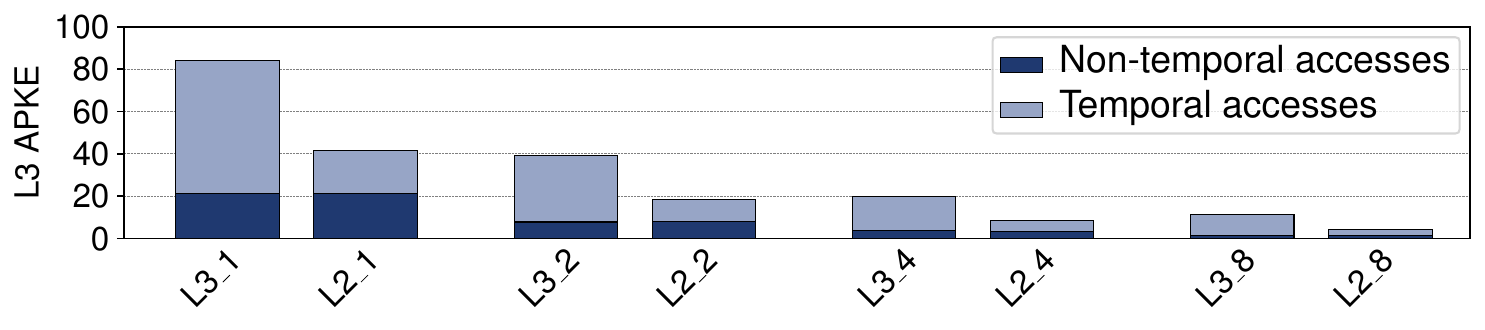}
    \vspace{-2em}
    \caption{L3 Accesses Per Kilo Element (APKE) of the BigBird gather function with different block sizes (1, 2, 4, and 8) and TMU configurations. Temporal accesses load indexes and non-temporal accesses load embedding vectors. Reading from L2 substantially helps filtering LLC accesses.}
    \label{fig:l3-accesses}
\end{figure}

As discussed in \Cref{sec:other-optimizations}, Ember lends itself to model-specific optimizations. \Cref{fig:l3-accesses} shows that, in block-sparse attention mechanisms (\Cref{sec:llms}), loading highly-reused embedding blocks from L2 rather than LLC filters 67\%---74\% of the embedding reads and 50\%---65\% of the overall accesses, which include non-temporal loads for indexes, with larger reductions on larger block sizes with more intrinsic reuse. By directly writing data with the store streams instead of going through the core, Ember enables efficient gather operations with low resource utilization.

\subsection{Comparison with Hand-optimized Code} \label{sec:hand-optimized-comparison} 

\begin{figure}
\includegraphics[width=\linewidth]{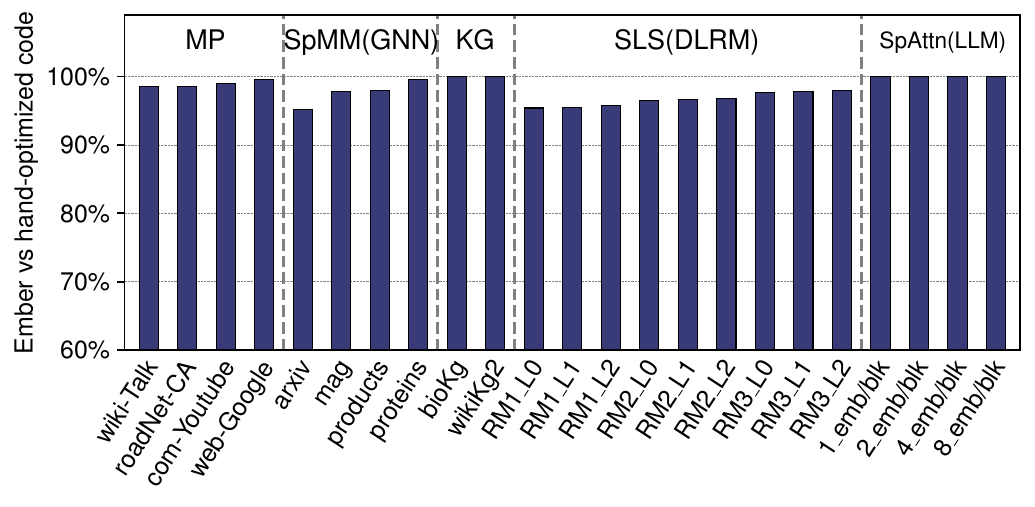} 
\vspace{-2.5em}
\caption{DAE code automatically generated and optimized by Ember (\texttt{emb-opt3}) achieves a geomean 99\% performance compared to hand-optimized code (\texttt{ref-dae}) across the different model classes in \Cref{tab:characterization}.}
\label{fig:ember-vs-handopt}
\end{figure}

\Cref{fig:ember-vs-handopt} shows the performance of DAE code automatically generated and optimized by Ember (\texttt{emb-opt3}) compared to hand-optimized code (\texttt{ref-dae}) for all model classes in \Cref{tab:characterization}. Besides all optimizations discussed in \Cref{sec:optimizations}, hand-optimized code also includes low-level, CPU-specific optimizations to improve callback invocations. These CPU-specific optimizations include, for instance, (1)~reordering the if-cases of multi-callback code (like the code in \Cref{fig:opt-impact-idx}) according to their taken frequency or (2)~set the values of control tokens to be directly used in compute code (e.g. to increment variables).

Overall, these CPU-specific optimizations primarily affect multi-callback code such as MP, SLS, and SpMM, yielding performance improvements of up to 5\%, with an average geometric mean improvement of 1\%. The limited impact of these low-level optimizations arises because, as discussed in \Cref{sec:ablation}, the optimizations introduced in \Cref{sec:optimizations} already push the architecture close to its limits, leaving little room for further gains. Nevertheless, because these low-level optimizations are highly CPU-specific, we chose not to integrate them into Ember, which is designed to provide a more general solution for a larger class of architectures.

Ultimately, we believe that the optimizations presented in \Cref{sec:optimizations} are sufficient to fully unlock the potential of general DAE architectures at no programmability cost.

%% file: sec_09_related.tex
\section{Related Work} \label{sec:related-work}

Embedding operations are becoming increasingly critical in several machine learning models, and DAE architectures a more widely adopted solution for similar irregular workloads. However, no study demonstrates the potential of DAE architectures on such a large class of models, and no compiler can generate DAE code that matches the performance of hand-optimized DAE embedding operations, as explained in this section.

\paragraph{\textbf{Embedding operations in Machine Learning Compilers:}}
For traditional architectures such as CPUs and GPUs, end-users call embedding operations through multiple deep learning frameworks~\cite{tf2015whitepaper,pytorch2019neurips,caffe2014arxiv,jax2018github,pytorchgeometric2019github,glow2018arxiv}, which can either be hand-written or automatically generated from progressive lowering. However, to the best of our knowledge, no machine learning compiler integrates automated DAE compilation, especially for embedding operations. While accelerators such as Meta's MTIA~\cite{mtia2023isca} feature cores to accelerate embedding lookups, embedding operations for such an accelerator typically come as libraries written by experts in low-level code~\cite{mtia2023isca}. However, this solution does not scale to general embedding operations where the space of models~\cite{ogb2020neurips}, algorithmic variants~\cite{fusedmm2021ipdps}, algorithmic optimizations~\cite{graphsage2017hamilton}, and input formats~\cite{taco2017oopsla} is large. Furthermore, libraries prevent fusion across operations~\cite{jianjun2021gcnax}.

\paragraph{\textbf{Sparse Tensor Algebra Compilers:}} 
When interpreted as sparse tensor expressions, embedding operations can be compiled to CPUs~\cite{taco2017oopsla,mlirsparse2022taco,looplets2023cgo,mosaic2023pldi,indexstreams2023pldi}, GPUs~\cite{tacoscheduling2020oopsla,sparsetir2023asplos,liu2024unisparse}, FPGAs~\cite{liu2024unisparse}, and distributed systems~\cite{spdistal2022sc}, but not DAE architectures. The Sparse Abstract Machine (SAM)~\cite{sam2023asplos} compiles sparse tensor algebra to a stream dataflow representation, which might generate code for the access unit but not for the execute unit nor the marshaling code.

\paragraph{\textbf{DAE Programming Models in Other Domains:}}
Several authors already proposed different DAE architectures for various domains, each one with a different programming interface. Planar~\cite{planar2021acm} offloads strided and irregular memory accesses from CPUs to tiny near-memory cores. As Planar primarily targets scientific workloads, it can be used through software libraries (e.g. BLAS~\cite{lawson1979blas}) optimized by experts. SpZip~\cite{spzip2021isca}, instead, integrates near-core specialized units in multicore processors to accelerate traversal and (de)compression operations in graph analytics. As these algorithms are generally hand-written by end users, SpZip comes with its own domain-specific language. Finally, MAPLE~\cite{maple2022isca} offloads indirect accesses to specialized units placed in the network-on-chip of a multicore processor. As MAPLE targets general irregular workloads, it comes with its own compiler, DeSC~\cite{desc2015micro}, which decouples access from execute operations in LLVM~\cite{llvm2004cs}. A similar LLVM technique has been used to offload affine, indirect, and pointer-chasing memory accesses to stream engines on CPUs~\cite{stream2019wang}. However, the LLVM IR is an unstructured IR that represents loops with conditional branches. This substantially limits code analyses, decoupling, and optimizations of programs with complex loops like embedding operations~\cite{panda2020tcad}. While other frameworks, including HLS tools~\cite{panda2020tcad,cong2011autoesl}, proposed to overcome LLVM limitations by using pragmas~\cite{cad2022tc,dsagen2020weng}, this solution is not applicable to machine learning compilers. Moreover, the LLVM IR is not designed for DAE code, making advanced optimizations impractical.

%% file: sec_10_conclusions.tex
\section{Conclusions} \label{sec:conclusions}

In conclusion, we demonstrated that DAE architectures outperform GPUs by 2.6$\times$ in performance and 6.4$\times$ in performance/watt on embedding-intensive models. Then, we designed the Ember compiler to integrate DAE architectures in common machine-learning frameworks. Compared to other approaches in the literature, Ember progressively lowers embedding operations through custom intermediate representations to optimize code at different abstraction levels. In this way, Ember implements all the necessary optimizations, both local and global optimizations, to match the performance of hand-optimized code, enabling the potential of DAE architectures at no programmability cost.